\documentclass{article} % For LaTeX2e
\usepackage[final]{colm2026_conference}

\usepackage{microtype}
\usepackage{hyperref}
\usepackage{url}
\usepackage{booktabs}
\usepackage{lineno}
\usepackage{multicol}

\definecolor{darkblue}{rgb}{0, 0, 0.5}
\hypersetup{colorlinks=true, citecolor=darkblue, linkcolor=darkblue, urlcolor=darkblue}

\usepackage[T1]{fontenc}
\usepackage[utf8]{inputenc}
\usepackage{inconsolata}

\usepackage{graphicx}
\usepackage{harmony}
\usepackage[nointegrals]{wasysym}
\usepackage{musicography}
\usepackage{tcolorbox, enumitem}
\usepackage{array}
\usepackage{diagbox}
\usepackage{multirow}
\usepackage[table,dvipsnames]{xcolor}
\usepackage{stmaryrd}
\usepackage{CJKutf8}
\usepackage{amssymb}

\newcommand{\ssigmusraise}{-0.1ex}
\newcommand{\ssigmusscale}{0.85}
\newcommand{\ssigmus}[1]{\raisebox{\ssigmusraise}{\scalebox{\ssigmusscale}{#1}}}

\definecolor{light-pink}{HTML}{FFD1DC}
\definecolor{light-red}{HTML}{F2A6A6}
\definecolor{light-orange}{HTML}{FFE0B2}
\definecolor{light-yellow}{HTML}{FFF9C4}
\definecolor{light-green}{HTML}{C8E6C9}
\definecolor{light-cyan}{HTML}{B2EBF2}
\definecolor{light-blue}{HTML}{BBDEFB}
\definecolor{light-purple}{HTML}{E1BEE7}
\definecolor{gemini-blue}{RGB}{74, 144, 226}

\definecolor{lightlight-blue}{HTML}{E3F2FD}
\definecolor{light-gray}{RGB}{180, 180, 180}
\definecolor{lightlight-gray}{RGB}{235, 235, 235}

\definecolor{medium-pink}{HTML}{E91E63}
\definecolor{medium-orange}{HTML}{F57C00}
\definecolor{medium-yellow}{HTML}{FBC02D}
\definecolor{medium-green}{HTML}{388E3C}
\definecolor{medium-cyan}{HTML}{0097A7}
\definecolor{medium-blue}{HTML}{1976D2}
\definecolor{medium-purple}{HTML}{7B1FA2}

\newcommand{\cmark}{$\checkmark$}
\newcommand{\xmark}{$\times$}

\newcommand{\dddagger}{%
  \mathbin{\vphantom{\dagger}\smash{%
    \ooalign{%
      $\scalebox{0.8}{\(\dagger\)}$\cr
      \raise.5ex\hbox{\scalebox{0.8}{\(\dagger\)}}\cr
      \raise-.5ex\hbox{\scalebox{0.8}{\(\dagger\)}}\cr
    }%
  }}%
}

\newenvironment{memeexample}[1][]%
{\begin{center}\begin{minipage}{\linewidth}\begin{tcolorbox}[#1]}%
{\end{tcolorbox}\end{minipage}\end{center}}

\title{%
  AVMeme Exam: A Multimodal Multilingual Multicultural Benchmark for LLMs’ Contextual and Cultural Knowledge and Thinking  
}

\author{
\normalsize
\textbf{Xilin Jiang}\textsuperscript{$\dagger\sharp$},
\textbf{Qiaolin Wang}\textsuperscript{$\dagger\sharp$},
\textbf{Junkai Wu}\textsuperscript{$\dagger\natural$},
\textbf{Xiaomin He}\textsuperscript{$\dagger\sharp$},
\textbf{Zhongweiyang Xu}\textsuperscript{$\dagger\flat$}
\\\normalsize
\textbf{Yinghao Ma}\textsuperscript{$\ddagger$\,\ssigmus{\musQuarter}},
\textbf{Minshuo Piao}\textsuperscript{$\ddagger\sharp$},
\textbf{Kaiyi Yang}\textsuperscript{$\ddagger\flat$},
\textbf{Xiuwen Zheng}\textsuperscript{$\ddagger\flat$},
\textbf{Riki Shimizu}\textsuperscript{$\ddagger\sharp$}, 
\\\normalsize
\textbf{Yicong Chen}\textsuperscript{$\ddagger\natural$}
\textbf{Arsalan Firoozi}\textsuperscript{$\ddagger\sharp$},
\textbf{Gavin Mischler}\textsuperscript{$\ddagger\sharp$},
\textbf{Sukru Samet Dindar}\textsuperscript{$\ddagger\sharp$},
\\\normalsize
\textbf{Richard Antonello}\textsuperscript{$\ddagger\sharp$},
\textbf{Linyang He}\textsuperscript{$\ddagger\sharp$},
\textbf{Tsun-An Hsieh}\textsuperscript{$\ddagger\flat$}
\textbf{Xulin Fan}\textsuperscript{$\ddagger\flat$},
\textbf{Yulun Wu}\textsuperscript{$\ddagger\flat$},
\\\normalsize
\textbf{Yuesheng Ma}\textsuperscript{$\ddagger\sharp$},
\textbf{Chaitanya Amballa}\textsuperscript{$\ddagger\flat$},
\textbf{Weixiong Chen}\textsuperscript{$\ddagger$\,\ssigmus{\musQuarter}},
\textbf{Jiarui Hai}\textsuperscript{$\ddagger$\,\ssigmus{\musSixteenth}},
\textbf{Ruisi Li}\textsuperscript{$\ddagger$\,\ssigmus{\musEighth}},
\\\normalsize
\textbf{Vishal Choudhari}\textsuperscript{$\ddagger\sharp$},
\textbf{Cong Han}\textsuperscript{$\ddagger$\,\musHalf},
\textbf{Yinghao Aaron Li}\textsuperscript{$\ddagger$\,\musWhole},
\\\normalsize
\textbf{Adeen Flinker}\textsuperscript{$\dddagger$\,\ssigmus{\musEighth}},
\textbf{Mounya Elhilali}\textsuperscript{$\ddagger$\,\ssigmus{\musSixteenth}},
\textbf{Emmanouil Benetos}\textsuperscript{$\dddagger$\,\ssigmus{\musQuarter}},
\\ \normalsize
\textbf{Mark Hasegawa-Johnson}\textsuperscript{$\dddagger\flat$},
\textbf{Romit Roy Choudhury}\textsuperscript{$\dddagger\flat$},
\textbf{Nima Mesgarani}\textsuperscript{$\dddagger\sharp$}
\\[4pt]
\small
\textsuperscript{$\sharp$}Columbia University,\
\textsuperscript{$\flat$}University of Illinois Urbana\mbox{-}Champaign,\
\textsuperscript{$\natural$}University of Washington,
\\[-2pt]\small
\textsuperscript{\ssigmus{\musSixteenth}}Johns Hopkins University,\
\textsuperscript{\ssigmus{\musEighth}}New York University,\
\textsuperscript{\ssigmus{\musQuarter}}Queen Mary University of London,
\\[-2pt]\small
\textsuperscript{\musHalf}Google,\
\textsuperscript{\musWhole}Meta
\\[3pt]
\fontsize{7.5pt}{9pt}\selectfont
\{xj2289, qw2443, xh2369\}@columbia.edu,\
junkaiwu@uw.edu,\
zx21@illinois.edu,\
nima@ee.columbia.edu
}

\begin{document}

\ifcolmsubmission
\linenumbers
\fi

\maketitle

% \begingroup
%   \renewcommand{\thefootnote}{}
%   \footnotetext{\hspace{-0.6cm}$\dagger$ Co-first; $\ddagger$ Co-second; $\dddagger$ Professors. Q.W., J.W., Z.X.: LLM evaluation; X.H.: human evaluation; X.J.: project lead.}
% \endgroup

\begin{abstract}
Internet audio-visual clips convey meaning through time-varying sound and motion, which extend beyond what text alone can represent. To examine whether AI models can understand such signals in human cultural contexts, we introduce AVMeme Exam, a human-curated benchmark of over one thousand iconic Internet sounds and videos spanning speech, songs, music, and sound effects. Each meme is paired with a unique Q\&A assessing levels of understanding from surface content to context and emotion to usage and world knowledge, along with metadata such as original year, transcript, summary, and sensitivity. We systematically evaluate state-of-the-art multimodal large language models (MLLMs) alongside human participants using this benchmark. Our results reveal a consistent limitation: current models perform poorly on textless music and sound effects, and struggle to think in context and in culture compared to surface content. These findings highlight a key gap in human-aligned multimodal intelligence and call for models that can perceive contextually and culturally beyond the surface of what they hear and see. \\
Project page: avmemeexam.github.io/review
\end{abstract}

\begin{figure}[h]
    \centering
    \vspace{-0.4cm}
    \includegraphics[width=0.75\linewidth]{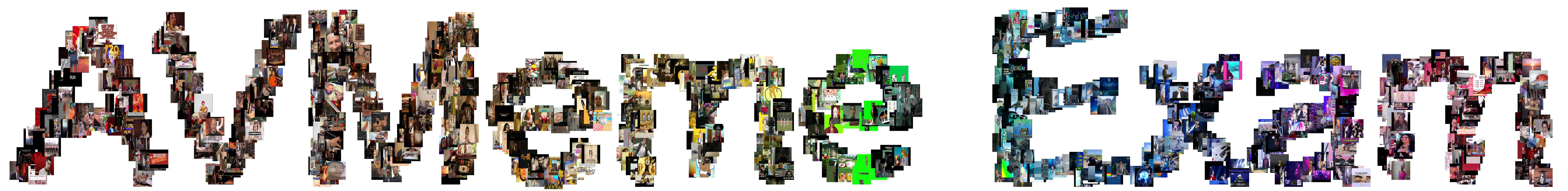}
    \caption{A mosaic art of ``AVMeme Exam" dotted by video frames from the meme clips.}
\end{figure}

% after
\begin{figure*}[h]
    \centering
    \includegraphics[width=0.9\linewidth]{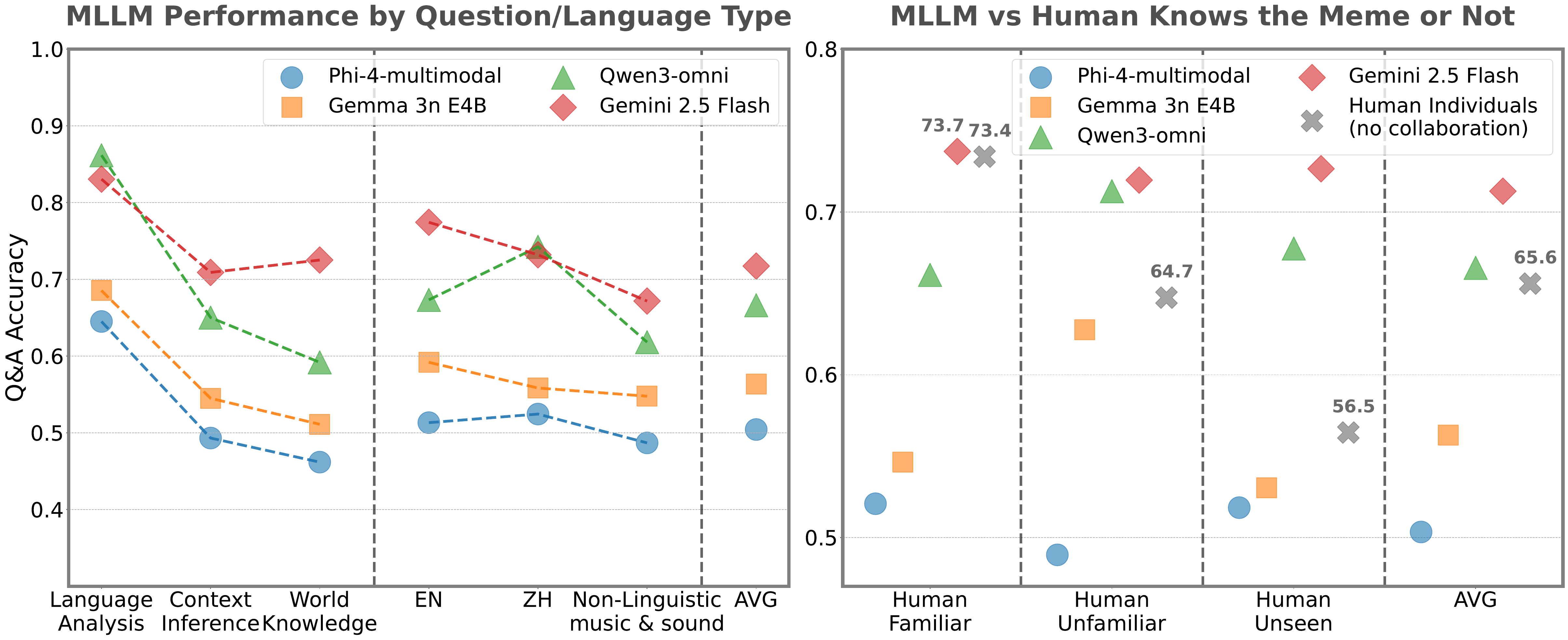}
    \caption{AVMeme Exam tests multimodal understanding \underline{beyond} language and surface content. \textbf{Left:} Multimodal LLMs understand content much better than context, culture, and world knowledge, and media clips with spoken language than those without. \textbf{Right:} Open-source LLMs still lose to humans on humans' own familiar memes, while humans remain robust even on their unseen clips, showing that current LLMs are still limited in contextual and cultural thinking and real-world multimedia experience than real humans.}
    \label{fig:human_vs_llm}
\end{figure*}

\section{Introduction}

Artificial General Intelligence (AGI) \cite{legg2007universal, morris2023levels, bubeck2023sparks, balepur-etal-2025-best} demands more than mastery of words. Human perception and communication also rely on sound and vision, along with other sensory inputs. AGI should also understand not only what is said or shown on the surface, but what is actually being communicated: the sarcasm in a voice, the emotion in a song, the sense of triumph or defeat in wordless music, and how billions of people can instantly recognize a second-long sound cue — 
from \textit{Beethoven's fate motif} \
\(
\textcolor{medium-green}{\mathrm{G}}{-}%
\textcolor{medium-green}{\mathrm{G}}{-}%
\textcolor{medium-green}{\mathrm{G}}{-}%
\textcolor{medium-pink}{\mathrm{E}^\flat}
\) 
to \textit{Nokia ringtone} 
\(
\textcolor{medium-purple}{\mathrm{E2}}{-}%
\textcolor{medium-blue}{\mathrm{D2}}{-}%
\textcolor{medium-orange}{\mathrm{F}^\sharp1}{-}%
\textcolor{medium-cyan}{\mathrm{G}^\sharp1}
\). Yet today's multimodal large language models (MLLMs) \cite{alayrac2022flamingo, liu2023visual, openai2024gpt4o, gong2024listen, tang2024salmonn,yin2024survey} still face key challenges. First, time-varying audio and video signals convey meaning through prosody, melody, pacing, etc, all of which cannot be fully described by language. Second, understanding surface content, who says what or what happens, is only a starting point. Models must also understand why a moment is recognizable, how it is used, and what cultural significance it carries.

To examine whether MLLMs can serve as empathetic assistants that can resonate with human users, we study a pervasive yet understudied medium of online communication: the audio-visual \textit{meme}. We define memes broadly as recognizable audio-visual clips (not necessarily humorous), including movie lines, sound effects, and musical phrases that people reuse with stable communicative purposes to express emotions and intentions. Because their meaning relies on delivery and shared culture besides content itself, they are an effective testbed for multimedia world understanding. Our central question is: \textit{To what extent can MLLMs understand audio-visual memes including their literal content, underlying context, emotion, usage, and cultural grounding. Where do they succeed or fail?}

Yet, no existing audio-visual benchmark dives into aspects of context, culture, usage, and world knowledge. Audio benchmarks such as Dynamic-SUPERB \cite{dynamic-superb}, AudioBench \cite{audiobench}, MMAU \cite{mmau}, ADU-Bench \cite{adu-bench}, and MMAR \cite{mmar} focus on recognition, captioning, dialogue understanding, and acoustic and semantic reasoning. While challenging, answers still reside in the waveform itself. Audio-visual benchmarks such as AV-SUPERB \cite{av-superb}, AV-Odyssey \cite{av-odyssey}, DAVE \cite{DAVE}, AV-SpeakerBench \cite{AV-SpeakerBench}, and AVQA variants \cite{music-AVQA, AVQA} examine event detection, ordering, alignment, and causality, but are still limited to what appears within the video frames. Meanwhile, a few language and image meme benchmarks indeed explore contextual and cultural implications, including Hateful Memes Challenge \cite{hateful_memes}, What-Do-You-Meme \cite{what_do_you_meme}, GOAT-Bench \cite{Goat-Bench}, MemeSafetyBench \cite{Meme-SafetyBench}, MemeReaCon \cite{MemeReaCon}. But unlike static text and image memes, the time dimension of audio-visual memes carries rich temporal dynamics, emotional shifts, and other culturally meaningful cues that prior meme and cultural benchmarks have not studied.

\begin{figure*}[t]
    \centering
    \includegraphics[width=\linewidth]{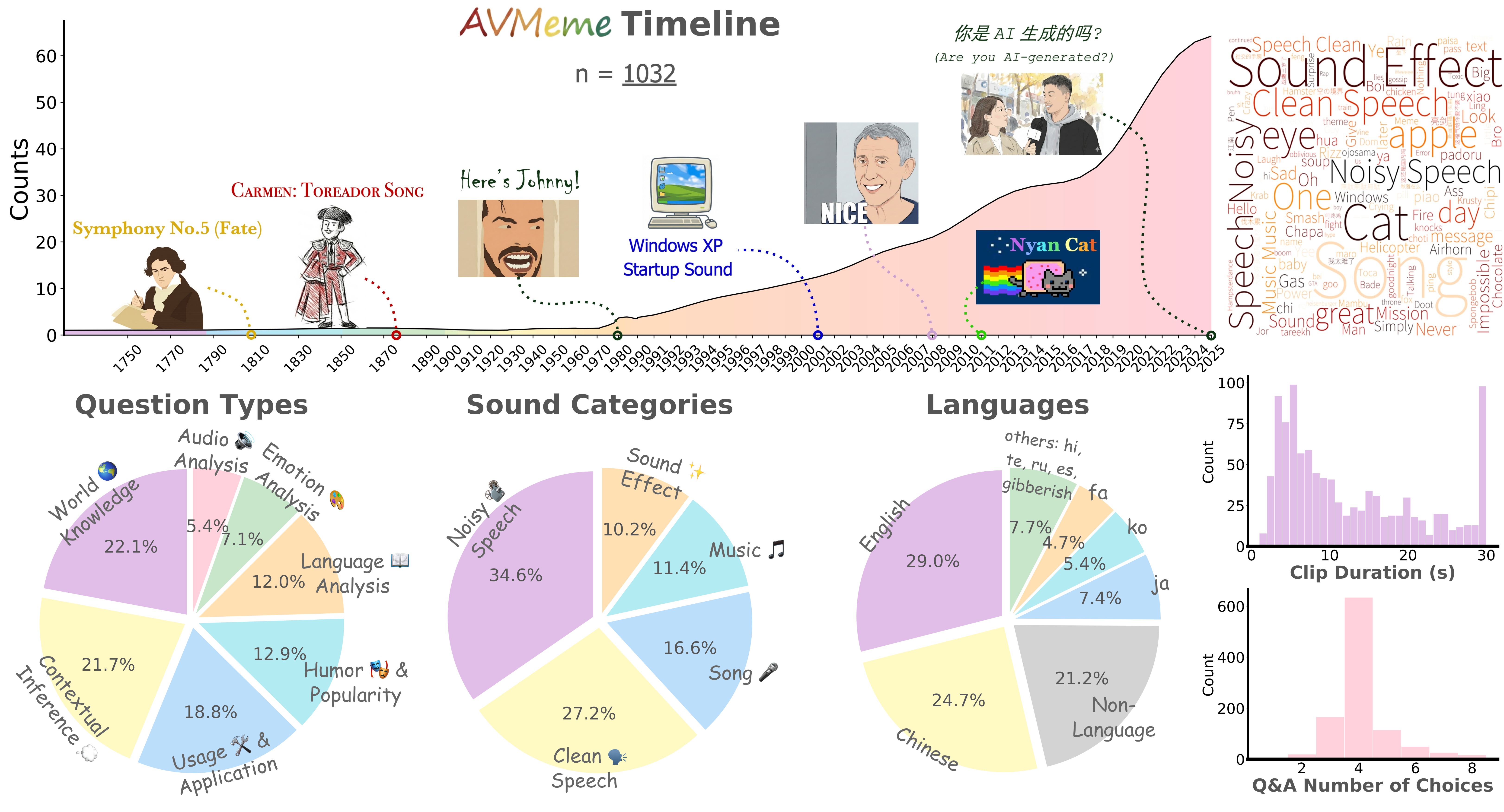}
    \caption{\textbf{Top:} Historical timeline of the 1,032 audio-visual memes curated in AVMeme Exam, spotlighting famous music rhythm, movie lines, sound effects, and viral Internet memes. \textbf{Bottom:} Pie charts summarize the distributions of question types, sound categories, and languages, highlighting the data diversity. \textbf{Right:} Frequent words in the memes' names and distributions of clip durations and number of choices.  The duration is cut to 30 seconds, which is the maximum input audio length for most models.}  
    \label{fig:meme_stat}
\end{figure*}

To bridge these gaps, we introduce \textbf{AVMeme Exam}, a manual collection of over one thousand iconic Internet sounds and videos annotated with human-annotated metadata and human-written questions probing surface understanding, contextual inference, emotion, humor, usage, and world knowledge. Evaluation of state-of-the-art MLLMs and human participants reveals consistent limitations in models’ ability to recognize, interpret, and culturally situate such clips. Together, our contributions establish AVMeme Exam as a comprehensive resource for diagnosing contextual and cultural weaknesses of AIs and guiding future progress in human-aligned multimodal intelligence.

We will first describe the AVMeme Exam in Section \ref{sec:dataset}, and then present and analyze model performance in Section \ref{sec:results}, along with key insights.

\section{AVMeme Exam}
\label{sec:dataset}

\begin{figure*}[t]
    \centering
    \includegraphics[width=\linewidth]{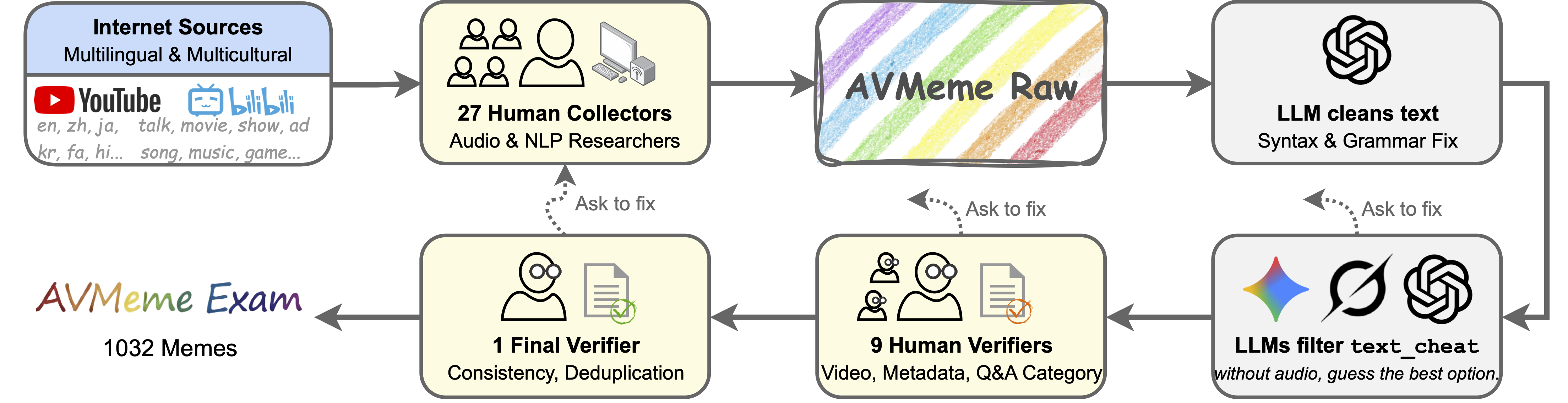}
    \caption{AVMeme Exam collection \& verification pipeline. Videos and Q\&As are human collected and verified (yellow). LLMs (gray) are used for text cleanup and to detect questions easily answered by text without audio given.}  
    \label{fig:data_pipeline}
    \vspace{-0.7cm}
\end{figure*}

Figure \ref{fig:data_pipeline} illustrates the overall collection and verification pipeline of AVMeme Exam. More details are documented in Appendix \ref{app:collection}.

\subsection{Collection}
The collection of AVMeme Exam is guided by three pillars: \textcolor{medium-pink}{\textbf{Human-collected}}: all clips are selected and annotated by 27 audio and NLP researchers, raised in the U.S., China, Japan, India, Middle East, and etc. They personally recognize and use these memes, ensuring that each entry is chosen with human judgment rather than automatic web scraping. \textcolor{medium-green}{\textbf{Audio-centric}}: sound serves as the primary media of meaning, complemented by visual, spanning speech, songs, music, and wordless sound effects. \textcolor{medium-blue}{\textbf{Multicultural-grounded}}: the diverse linguistic and cultural backgrounds of contributors enables coverage of memes originating from East and South Asia, Middle East, Europe, and North America. Collectively, these pillars ensure that AVMeme Exam reflects authentic, culturally embedded, real-world multimodal communication.

\begin{figure*}[t]
\centering

\begin{minipage}[t]{0.49\linewidth}
\begin{tcolorbox}[
  colback=gray!2,
  colframe=light-red,
  boxrule=1.2pt,
  title={You shall not pass},
  left=4pt,right=4pt,top=4pt,bottom=4pt
]
\small
\color{black!70}

\noindent
\begin{minipage}[t]{0.6\linewidth}
    \vspace{0pt}
    \textbf{\textcolor{black}{Category:}} noisy speech \\
    \textbf{\textcolor{black}{Year:}} 2001 \\
    \textbf{\textcolor{black}{Language:}} English \\
\end{minipage}%
\hfill
\begin{minipage}[t]{0.35\linewidth}
    \vspace{0pt}
    \raggedleft
    \includegraphics[width=\linewidth]{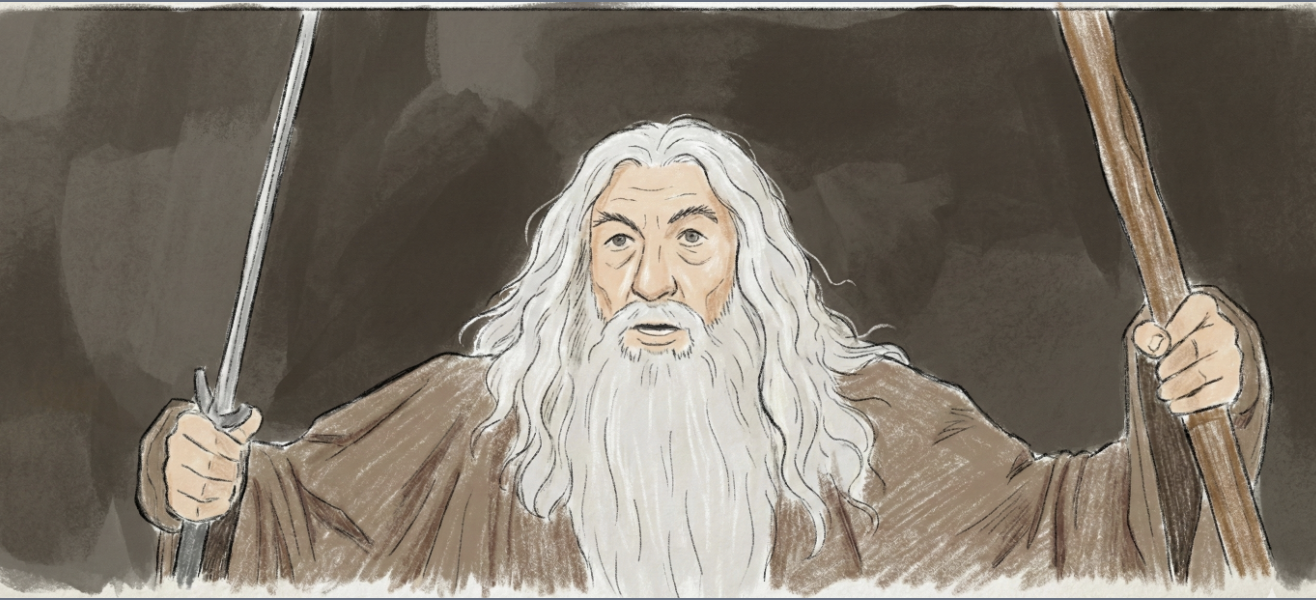}
\end{minipage}

\par\vspace{4pt}

\textbf{\textcolor{black}{Transcription:}} You shall not pass. \\
\textbf{\textcolor{black}{Summary:}}
A powerful wizard confronts a fiery monster on a narrow bridge, striking his staff and shouting defiantly, "You shall not pass!". \\[3pt]
\textbf{\textcolor{black}{Usage:}}
Used to jokingly or seriously deny someone access or prevent them from proceeding. \\
\textbf{\textcolor{black}{Emotion:}}
fearful/anxious \\
\textbf{\textcolor{black}{Sensitivity:}}
violence/gore \\
\textbf{\textcolor{black}{Question:}}
\emph{Which of the following is a humorous way a college student might use this meme?}
\begin{enumerate}[label=\textbf{\Alph*.}, nosep, leftmargin=*]
    \item When talking about the spread of the flu among students.
    \item \textbf{When discussing how difficult the final exam was in a class.}
    \item When presenting a final project in an art class.
    \item When asking the professor a question during a lecture.
\end{enumerate}
\vspace{3pt}
\textbf{\textcolor{black}{Question Type:}} Usage \& Application
\end{tcolorbox}
\end{minipage}
\hfill
\begin{minipage}[t]{0.49\linewidth}
\begin{tcolorbox}[
  colback=gray!2,
  colframe=light-orange,
  boxrule=1.2pt,
  title={Never Gonna Give You Up},
  left=4pt,right=4pt,top=4pt,bottom=4pt
]
\small
\color{black!70}

\noindent
\begin{minipage}[t]{0.6\linewidth}
    \vspace{0pt}
    \textbf{\textcolor{black}{Category:}} song \\
    \textbf{\textcolor{black}{Year:}} 1987 \\
    \textbf{\textcolor{black}{Language:}} English \\
    \textbf{\textcolor{black}{Transcription:}} We're no strangers to love. You know the rules and...
\end{minipage}%
\hfill
\begin{minipage}[t]{0.3\linewidth}
    \vspace{0pt}
    \raggedleft
    \includegraphics[width=\linewidth]{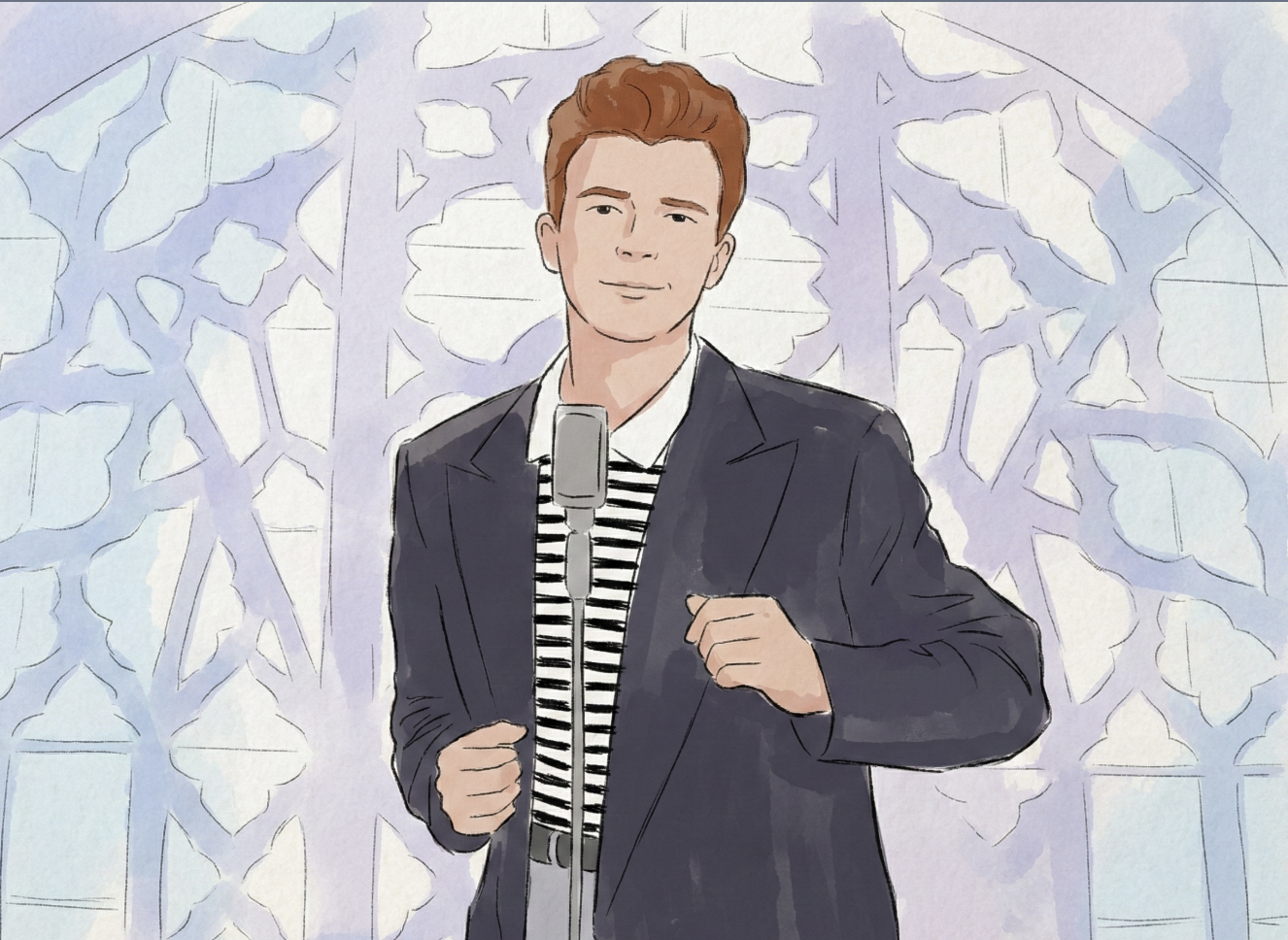}
\end{minipage}

\par\vspace{4pt}

\textbf{\textcolor{black}{Summary:}}
The start of music video "Never Gonna Give You Up". \\[3pt]
\textbf{\textcolor{black}{Usage:}}
A bait and switch is usually done using a disguised hyperlink leading to this music video. \\
\textbf{\textcolor{black}{Emotion:}}
surprised/shocked, nostalgic \\
\textbf{\textcolor{black}{Sensitivity:}}
none \\
\textbf{\textcolor{black}{Question:}}
\emph{What is the top reason this song became so popular as a meme?}
\begin{enumerate}[label=\textbf{\Alph*.}, nosep, leftmargin=*]
    \item \textbf{Unexpected appearance on the Internet.}
    \item Often got edited and remixed for parody.
    \item A lot of references from other songs and movies.
    \item The singer has sung it a lot.
    \item It became a sports anthem.
    \item It is popularized from forums such as 4chan and reddit.
\end{enumerate}
\vspace{3pt}
\textbf{\textcolor{black}{Question Type:}} Humor \& Popularity
\end{tcolorbox}
\end{minipage}

\caption{Meme examples. More examples are provided in Appendix \ref{app:examples}. }
\label{fig:avmeme-examples}
\vspace{-0.7cm}
\end{figure*}

In total, we collect \textbf{1,032} audio-visual memes spanning more than ten languages and five sound categories, with their distributions shown in Figure~\ref{fig:meme_stat}. Each meme consists of a video segment (URL, onset, offset) sourced from YouTube (86.1\%) and Bilibili (13.9\%), most exceeding one million views. It includes human-annotated metadata (summary, transcript, year, etc) and a multiple-choice Q\&A. To maximize diversity, we place no restrictions on the videos (topic, region, etc), but safeguards still apply. We prohibit political materials and explicit depictions of sexual, violent, hateful, criminal, or drug-related content. Implicit depictions, verbal or contextual, are annotated using a sensitivity attribute drawn from \{\textit{sex, violence/gore, drug/alcohol, race/gender/geography/identity, disadvantaged/disability}\}. Despite these safeguards, \textbf{\textit{some clips may remain sensitive for certain people, and discretion is advised.}} The emotion attribute takes one or multiple from \{\textit{happy, sad, angry/annoyed, fearful/anxious, surprised/shocked, sarcastic/ironic, cringe/embarrassed, nostalgic, neutral}\}. The question type is one of the seven (elaborated later). Figure~\ref{fig:avmeme-examples} shows two examples.

Finally, we prompt GPT-5.1 \cite{gpt5.1} for syntax and grammar fix of the summary, usage, and Q\&A fields, \textit{``Do minimal or no change if possible. Try not to add or change words unless necessary."} to preserve the human annotations as original as possible.

\subsection{Verification}

Nine human verifiers check every entry, reviewing all metadata fields: transcription, summary, usage, sensitivity, language, video integrity, and the multiple-choice Q\&A, on a clip-by-clip basis, and return to contributors to fix if any issues are identified.

While this process ensures syntactic correctness and annotation quality, it is not sufficient to guarantee that the Q\&A genuinely tests multimodal understanding, as some questions may still be solvable through unintended shortcuts \cite{zheng2024large, balepur-etal-2025-best}. We therefore begin with \textcolor{medium-pink}{\textbf{text-cheat}} detection. We run three LLMs in text-only modes: Gemini 2.5 Flash \cite{gemini2.5}, Grok~4 \cite{grok4}, and GPT-5.1 \cite{gpt5.1}, with prompt “\textit{without audio, guess the best option}.” Q\&As that all three models guess correctly are flagged as \texttt{text\_cheat}, for which the strong text priors allow the LLMs to infer the clip or the answer without listening or watching. For example, \textit{a famous sci-fi movie from the late 1970s} almost inevitably points to \textit{Star Wars}. Aware of this problem, we explicitly prohibit contributors from mentioning names, authors, historical events, or other giveaway keywords in the Q\&A. 
A small number of Q\&As correctly guessed by all three LLMs are still retained, since they may be guessed by chance or by choosing the most well-known option. For example, when \textit{John Williams}, the most famous film composer, appears among other composers, it is typically selected regardless of question phrasing. The full dataset is denoted \texttt{meme-full} (1032 memes), and removing all \texttt{text\_cheat} items yields \texttt{meme-main} (846 memes). From \texttt{meme-full} to \texttt{meme-main}, the deaf-and-blind guessing accuracy drops from \underline{34.7}$\rightarrow$\underline{20.4}, \underline{37.9}$\rightarrow$\underline{24.3}, and \underline{40.9}$\rightarrow$\underline{27.9} for Gemini 2.5 Flash, Grok~4, and GPT-5.1, respectively, showing that \texttt{meme-main} represents a more exclusive test of multimodal understanding.

Next, we assess \textcolor{medium-pink}{\textbf{visual-cheat}}. While we treat visual as a supplemental signal, it may trivially reveal the answer, especially through \textit{on-screen text}. We manually inspect each video and assign a \texttt{visual\_hint} attribute with one or multiple values: \{\textit{no text}, \textit{irrelevant text}, \textit{transcription}, \textit{title/name}, \textit{visual contains solution}\}. The last and the most severe category (labeled as \texttt{visual\_cheat}) denotes cases where the visual stream collapses the intended reasoning problem into simple OCR or object detection instead of genuine multimodal understanding. For \texttt{visual\_cheat} clips, we omit the visual input when evaluating audio–visual models.

These measures together ensure that we are evaluating MLLMs' genuine multimodal understanding abilities without text or visual shortcuts.

\begin{table*}[t]
\centering
\resizebox{\textwidth}{!}{
\setlength{\tabcolsep}{4pt}
\begin{tabular}{
p{0.2\textwidth}|
*{8}{>{\centering\arraybackslash}p{0.052\textwidth}}
|
*{8}{>{\centering\arraybackslash}p{0.052\textwidth}}
}
\toprule
\multirow{2}{*}{\diagbox[dir=SE]{Model}{QT}}
& \multicolumn{8}{c|}{\texttt{meme-full}}
& \multicolumn{8}{c}{\texttt{meme-main}} \\
\cmidrule(lr){2-9} \cmidrule(lr){10-17}
& \textcolor{medium-pink}{\textbf{A}} & \textcolor{medium-orange}{\textbf{L}}
& \textcolor{medium-yellow}{\textbf{C}} & \textcolor{medium-green}{\textbf{E}}
& \textcolor{medium-cyan}{\textbf{H}} & \textcolor{medium-blue}{\textbf{U}}
& \textcolor{medium-purple}{\textbf{W}} & \textbf{AVG}
& \textcolor{medium-pink}{\textbf{A}} & \textcolor{medium-orange}{\textbf{L}}
& \textcolor{medium-yellow}{\textbf{C}} & \textcolor{medium-green}{\textbf{E}}
& \textcolor{medium-cyan}{\textbf{H}} & \textcolor{medium-blue}{\textbf{U}}
& \textcolor{medium-purple}{\textbf{W}} & \textbf{AVG} \\
\midrule

%%%%%%%%%%%%%%%%%%%%%%%%%%%%%%%%%%%%%%%%%%%%%%%
%%%% I. AUDIO LLMs (sorted by year)
%%%%%%%%%%%%%%%%%%%%%%%%%%%%%%%%%%%%%%%%%%%%%%%

\multicolumn{17}{l}{\textbf{\large Audio LLMs}} \\
\hline

SALMONN
& 32.1 & 34.7 & 34.4 & 42.5 & 35.3 & 28.2 & 19.8 & 30.6
& 31.1 & 30.1 & 28.1 & 36.2 & 23.5 & 27.2 & 20.7 & 26.6 \\

Qwen2-Audio
& 51.8 & 52.0 & 41.7 & 49.3 & 46.6 & 41.8 & 25.1 & 41.0
& 42.2 & 47.6 & 35.9 & 44.8 & 28.2 & 36.6 & 22.1 & 34.2 \\

Audio Reasoner
& 39.3 & 42.6 & 33.2 & 47.9 & 38.2 & 38.1 & 23.5 & 35.1
& 35.6 & 44.6 & 29.0 & 39.7 & 32.1 & 34.8 & 23.7 & 32.1 \\

Kimi-Audio
& 50.0 & 60.5 & 47.8 & 61.6 & 52.6 & 49.2 & 35.2 & 48.5
& 44.4 & 56.3 & 41.6 & 55.2 & 31.8 & 44.4 & 33.7 & 42.1 \\

Audio Flamingo 3
& 46.4 & 62.9 & 47.8 & \textbf{65.8} & 55.6 & \textbf{53.8} & 28.6 & 48.7
& 40.0 & 58.3 & 40.5 & \textbf{62.1} & 36.5 & \textbf{48.1} & 26.4 & 41.7 \\

Step-Audio-2
& 48.2 & 34.7 & 42.9 & 38.4 & 52.6 & 42.1 & 26.4 & 39.3
& 37.8 & 24.3 & 34.1 & 27.6 & 31.8 & 35.2 & 24.5 & 30.3 \\

MiDashengLM
& 57.1 & 66.9 & 50.9 & 60.3 & \textbf{58.6} & 44.6 & 30.4 & 49.1
& \textbf{48.9} & 63.1 & 43.8 & 55.2 & \textbf{38.8} & 39.5 & 29.8 & 42.4 \\

MiMo-Audio
& \textbf{60.7} & 75.8 & 50.0 & 60.3 & 57.9 & 51.8 & 37.9 & 53.1
& 51.1 & 74.8 & 43.8 & 51.7 & \textbf{38.8} & 45.1 & 33.7 & 45.7 \\

Music Flamingo
& 55.4 & 63.7 & 50.9 & 64.4 & 55.6 & 51.3 & 31.3 & 50.0
& 44.4 & 58.3 & 43.2 & 58.6 & \textbf{38.8} & 46.3 & 28.8 & 42.8 \\

\hline
GPT-4o Audio $\dagger$
& 58.9 & \textbf{77.4} & \textbf{54.9} & \textbf{65.8} & 57.1 & 52.3 & \textbf{48.9} & \textbf{57.1}
& \textbf{48.9} & \textbf{75.7} & \textbf{46.5} & 60.3 & 37.6 & 46.9 & \textbf{45.6} & \textbf{50.1} \\

\midrule

%%%%%%%%%%%%%%%%%%%%%%%%%%%%%%%%%%%%%%%%%%%%%%%
%%%% II. AUDIO-VISUAL LLMs (filled)
%%%%%%%%%%%%%%%%%%%%%%%%%%%%%%%%%%%%%%%%%%%%%%%

\multicolumn{17}{l}{\textbf{\large Audio-Visual LLMs}} \\
\hline

Baichuan-Omni
& 38.5 & 50.0 & 45.2 & 52.8 & 47.4 & 38.7 & 26.0 & 40.9
& 31.7 & 43.7 & 37.6 & 45.6 & 27.1 & 32.7 & 23.0 & 33.1 \\
\rowcolor{lightlight-blue}
\hspace{2pt} + visual input
& 34.6 & 58.7 & 47.3 & 54.8 & 47.7 & 46.7 & 34.2 & 45.7
& 29.3 & 54.0 & 41.8 & 51.7 & 29.8 & 40.1 & 32.7 & 39.6 \\

MiniCPM
& 55.4 & 48.4 & 45.5 & 46.6 & 48.9 & 45.6 & 43.4 & 46.5
& 46.7 & 42.7 & 38.4 & 36.2 & 29.4 & 40.1 & 41.1 & 39.3 \\
\rowcolor{lightlight-blue}
\hspace{2pt} + visual input
& 57.1 & 53.2 & 45.1 & 56.2 & 51.1 & 49.0 & 44.2 & 48.8
& 48.9 & 47.6 & 36.8 & 48.3 & 28.2 & 45.3 & 42.0 & 41.6 \\

Phi-4-Multimodal
& 44.6 & 57.3 & 43.8 & 60.3 & 46.6 & 39.0 & 25.6 & 42.1
& 40.0 & 54.4 & 37.3 & 51.7 & 29.4 & 32.1 & 24.5 & 35.6 \\
\rowcolor{lightlight-blue}
\hspace{2pt} + visual input
& 42.9 & 62.1 & 48.4 & 61.6 & 48.5 & 41.8 & 31.4 & 45.8
& 33.3 & 55.3 & 41.8 & 55.2 & 28.6 & 34.2 & 29.9 & 38.3 \\

Qwen2.5-Omni
& 50.0 & 66.9 & 45.5 & 72.6 & 63.9 & 45.1 & 32.6 & 49.7
& 40.0 & 64.1 & 37.8 & 67.2 & 48.2 & 40.7 & 32.7 & 43.5 \\
\rowcolor{lightlight-blue}
\hspace{2pt} + visual input
& 62.5 & 85.4 & 62.3 & 75.3 & 63.9 & 59.3 & 45.3 & 62.0
& 53.3 & 83.3 & 58.7 & 70.7 & 48.2 & 55.3 & 43.1 & 56.7 \\

Gemma 3n E4B IT
& 46.4 & 61.3 & 47.8 & 56.2 & 56.4 & 40.0 & 23.3 & 44.2
& 35.6 & 56.3 & 42.7 & 48.3 & 36.5 & 34.0 & 19.7 & 36.4 \\
\rowcolor{lightlight-blue}
\hspace{2pt} + visual input
& 51.8 & 66.9 & 50.0 & 60.3 & 60.6 & 46.9 & 35.0 & 50.4
& 42.2 & 65.0 & 44.3 & 56.9 & 45.2 & 41.0 & 31.4 & 43.9 \\

Qwen3-Omni
& 64.3 & 79.0 & 59.6 & 75.3 & 65.4 & 59.5 & 45.8 & 61.0
& 55.6 & 76.7 & 56.5 & 69.0 & 50.6 & 54.3 & 42.8 & 55.4 \\
\rowcolor{lightlight-blue}
\hspace{2pt} + visual input
& 64.3 & 84.6 & 61.9 & 75.3 & 64.7 & 60.3 & 48.0 & 62.7
& 55.6 & 82.4 & 58.7 & 70.7 & 49.4 & 56.5 & 44.6 & 57.4 \\

\hline
Gemini 2.5 Flash $\dagger$
& 51.8 & 85.4 & 61.6 & 65.8 & 68.2 & 60.0 & 53.3 & 62.9
& 42.2 & 83.3 & 55.7 & 58.6 & 53.6 & 56.2 & 51.0 & 57.2 \\
\rowcolor{lightlight-blue}
\hspace{2pt} + visual input
& 56.4 & 82.3 & 69.2 & 78.1 & 72.2 & 60.6 & 63.6 & 68.3
& 47.7 & 79.6 & 66.1 & 72.4 & 60.0 & 57.1 & 62.1 & 63.9 \\
\rowcolor{lightlight-gray}
\hspace{2pt} \textit{- audio (text-only)}
& 37.5 & 33.9 & 36.8 & 31.5 & 48.9 & 35.4 & 25.1 & 34.8
& 22.2 & 20.4 & 23.4 & 13.8 & 20.0 & 22.2 & 18.3 & 20.5 \\

Gemini 3 Flash $\dagger$
& 64.3 & 86.3 & 75.4 & 75.3 & 77.4 & 67.7 & 67.4 & 73.2
& 55.6 & 84.5 & 71.4 & 72.4 & 67.1 & 64.2 & 66.3 & 69.1 \\
\rowcolor{lightlight-blue}
\hspace{2pt} + visual input
& 64.3 & 88.7 & 80.6 & 82.2 & \textbf{84.2} & \textbf{74.4} & 78.9 & 79.7
& 55.6 & 87.4 & 78.8 & 79.3 & \textbf{76.5} & 70.4 & 77.9 & 76.6 \\

Gemini 3 Pro $\dagger$
& 76.8 & \textbf{94.4} & 77.8 & 80.8 & 78.2 & 73.7 & 78.9 & 79.5
& 71.1 & \textbf{94.2} & 74.7 & 77.6 & 70.6 & 70.2 & 77.9 & 76.6 \\
\rowcolor{lightlight-blue}
\hspace{2pt} + visual input
& \textbf{80.4} & 91.1 & \textbf{81.1} & \textbf{86.3} & 83.5 & 72.8 & \textbf{83.6} & \textbf{81.9}
& \textbf{75.6} & 90.3 & \textbf{80.4} & \textbf{84.5} & 75.3 & \textbf{71.0} & \textbf{83.1} & \textbf{80.0} \\
\bottomrule
\end{tabular}
}
\caption{Model performance across question types on \texttt{meme-full} and \texttt{meme-main}.}
\label{tab:main_QT}
\vspace{-0.5cm}
\end{table*}

\subsection{Question Types}

Q\&As are categorized into seven question types by human verifiers, ensuring consistent labeling while allowing contributors to draft Q\&As creatively without prior constraints.

\vspace{2pt}
\noindent
\textcolor{black!60}{$\ulcorner$The first two types ask for the immediate information of the clip: what is heard or said. It reflects the most basic level of multimodal comprehension.$\lrcorner$}
\vspace{2pt}

\noindent
\textbf{\textcolor{medium-pink}{A}udio Analysis} focuses on what the sound alone reveals: prosody, rhythm, style, or other audible patterns. They check whether a model can genuinely analyze audio rather than rely on the text. \underline{Example}: \textit{What kind of musical processing was applied to the human voice?}
%\textit{How does the man sound in the clip?} and 

\noindent
\textbf{\textcolor{medium-orange}{L}anguage Analysis} tests recognition and parsing of the spoken words, including what is said and how it functions within a sentence or conversation. The answer comes from the linguistic content of the clip. \underline{Example}: \textit{What one thing does the speaker claim about himself?} %and \textit{Who is or isn't causing the problem according to the speaker?}

\vspace{2pt}
\noindent
\textcolor{black!60}{$\ulcorner$The middle three types move beyond literal content to how humans grasp meaning, feeling, and humor in a clip. It often requires educated thinking.$\lrcorner$} 

\noindent
\textbf{\textcolor{medium-yellow}{C}ontextual Inference} evaluates whether a model can understand the situation behind the clip: what the speaker intends, what they mean, or what is happening in the scene. The correct answer comes from interpreting the context, not just rephrasing the words. \underline{Example}: \textit{Which situation best matches the tone and intent of this moment?} %and \textit{According to the man's logic, what is true about ice coke?}

\noindent
\textbf{\textcolor{medium-green}{E}motion Analysis} asks the model to identify the feeling expressed, based on tone, delivery, pacing, or the effect on the audience. They check whether the model can pick up how the moment is meant to feel. \underline{Example}: 
\textit{What will people likely feel when they first listen to this piece?}
%\textit{Which emotional mix best describes the way this line is spoken?} and \textit{What will people likely feel when they first listen to this piece?}

\noindent
\textbf{\textcolor{medium-cyan}{H}umor \& Popularity} asks why a clip became funny, iconic, or widely shared. They involve noticing unexpected reactions, exaggerated behavior, or other traits that made the moment memorable online. \underline{Example}: \textit{Which of the following is not a reason that the clip is humorous?} %and \textit{Which part of this speech became the most famous?} 

\begin{table*}[t]
\centering
\resizebox{\textwidth}{!}{
\setlength{\tabcolsep}{4pt}
\begin{tabular}{
p{0.22\textwidth}|
*{4}{>{\centering\arraybackslash}p{0.036\textwidth}}
|
*{4}{>{\centering\arraybackslash}p{0.036\textwidth}}
|
*{6}{>{\centering\arraybackslash}p{0.036\textwidth}}
|
*{6}{>{\centering\arraybackslash}p{0.036\textwidth}}
}
\toprule
\multirow{2}{*}{\diagbox[dir=SE]{Model}{Category}}
& \multicolumn{4}{c|}{\texttt{meme-full}: ST}
& \multicolumn{4}{c|}{\texttt{meme-main}: ST}
& \multicolumn{6}{c|}{\texttt{meme-full}: Lang}
& \multicolumn{6}{c}{\texttt{meme-main}: Lang} \\
\cmidrule(lr){2-5}
\cmidrule(lr){6-9}
\cmidrule(lr){10-15}
\cmidrule(lr){16-21}
& Sp & So & Mu & Sfx
& Sp & So & Mu & Sfx
& en & zh & ja & ko & fa & nan
& en & zh & ja & ko & fa & nan \\
\midrule
\multicolumn{21}{l}{\textbf{Audio LLMs}} \\
\hline

SALMONN
& 32.6 & 25.1 & 25.4 & 33.3 & 26.7 & 22.8 & 25.5 & 33.0 & 30.0 & 32.3 & 25.0 & 35.7 & 35.4 & 27.2 & 26.7 & 23.8 & 20.3 & 36.7 & 26.8 & 27.5 \\

Qwen2-Audio
& 43.5 & 40.8 & 37.3 & 30.5 & 35.9 & 34.7 & 32.4 & 26.8 & 40.1 & 51.1 & 33.8 & 51.8 & 29.2 & 33.6 & 34.3 & 39.7 & 30.2 & 51.0 & 26.8 & 29.0 \\

Audio Reasoner
& 36.6 & 35.1 & 33.9 & 27.6 & 32.7 & 34.5 & 30.4 & 26.8 & 35.8 & 37.2 & 32.9 & 41.1 & 31.9 & 29.5 & 34.1 & 31.2 & 29.7 & 38.8 & 32.5 & 26.9 \\

Kimi-Audio
& 51.3 & 48.0 & 42.4 & \textbf{40.0} & 42.8 & 44.1 & 38.2 & \textbf{39.2} & 51.3 & 50.8 & 42.1 & 46.4 & \textbf{47.9} & 39.6 & 46.3 & 37.0 & 40.6 & 42.9 & \textbf{46.3} & 36.3 \\

Audio Flamingo 3
& 53.1 & 45.6 & 43.2 & 33.3 & 44.6 & 41.4 & \textbf{40.2} & 28.9 & 50.0 & 54.1 & 43.4 & 51.8 & 35.4 & 40.1 & 45.5 & 40.7 & 37.5 & 46.9 & 34.1 & 35.8 \\

Step-Audio-2
& 42.5 & 38.6 & 28.8 & 33.3 & 31.3 & 32.4 & 22.5 & 29.9 & 39.3 & 42.9 & 35.5 & 37.5 & 43.8 & 32.3 & 32.9 & 24.3 & 28.1 & 32.7 & 36.6 & 27.5 \\

MiDashengLM
& 54.4 & 44.4 & 39.0 & 36.2 & 46.2 & 40.0 & 35.3 & 34.0 & 51.0 & 57.9 & 31.6 & 53.6 & 41.7 & 38.2 & 46.3 & 44.4 & 28.1 & 49.0 & 41.5 & 35.2 \\

MiMo-Audio
& 58.0 & 52.6 & \textbf{44.1} & 34.3 & 49.8 & 48.3 & 36.3 & 30.9 & 57.0 & \textbf{60.2} & 46.1 & 48.2 & 39.6 & 40.1 & 52.2 & \textbf{47.6} & 39.1 & 44.9 & 39.0 & 33.7 \\

Music Flamingo
& 55.5 & 48.5 & 35.6 & 35.2 & 46.8 & 43.4 & 32.4 & 32.0 & 51.0 & 57.9 & 48.7 & \textbf{60.7} & 39.6 & 35.5 & 47.1 & 44.4 & 40.6 & \textbf{57.1} & 36.6 & 31.6 \\

\hline
GPT-4o Audio \dag
& \textbf{62.5} & \textbf{59.4} & 42.4 & 37.1 & \textbf{54.5} & \textbf{55.6} & 36.3 & 34.0 & \textbf{65.0} & 59.0 & \textbf{57.3} & 57.1 & 41.7 & \textbf{41.2} & \textbf{60.0} & 46.0 & \textbf{52.4} & 55.1 & 34.1 & \textbf{36.5} \\

\midrule
\multicolumn{21}{l}{\textbf{\large Audio-Visual LLMs}} \\
\hline

Baichuan-Omni
& 45.7 & 38.6 & 28.4 & 29.0 & 36.8 & 32.9 & 22.0 & 25.0 & 40.7 & 51.3 & 40.3 & 40.0 & 38.3 & 28.3 & 34.6 & 38.5 & 34.4 & 41.7 & 31.7 & 22.9 \\
\rowcolor{lightlight-blue}
\hspace{2pt} + visual input
& 49.3 & 45.3 & 35.9 & 35.0 & 41.8 & 42.4 & 30.7 & 33.3 & 47.9 & 50.4 & 42.7 & 50.0 & 34.0 & 35.4 & 44.4 & 37.6 & 39.1 & 51.0 & 32.5 & 31.2 \\

MiniCPM
& 48.6 & 42.7 & 44.4 & 41.9 & 39.2 & 37.2 & 42.6 & 39.2 & 44.0 & 50.8 & 39.5 & 35.7 & 39.6 & 46.8 & 38.4 & 36.5 & 32.8 & 30.6 & 36.6 & 44.8 \\
\rowcolor{lightlight-blue}
\hspace{2pt} + visual input
& 51.6 & 43.9 & 45.7 & 43.8 & 42.8 & 37.2 & 43.0 & 40.2 & 49.0 & 52.6 & 38.2 & 41.1 & 41.7 & 47.0 & 43.9 & 37.6 & 31.2 & 38.8 & 39.0 & 44.0 \\

Phi-4-Multimodal
& 47.3 & 37.4 & 31.4 & 29.5 & 40.0 & 32.4 & 26.5 & 26.8 & 45.0 & 48.5 & 39.5 & 42.9 & 33.3 & 30.9 & 40.8 & 36.0 & 34.4 & 40.8 & 34.1 & 26.4 \\
\rowcolor{lightlight-blue}
\hspace{2pt} + visual input
& 51.0 & 40.4 & 35.0 & 34.3 & 41.9 & 36.6 & 31.7 & 28.7 & 46.7 & 51.7 & 42.1 & 39.3 & 42.6 & 34.3 & 43.1 & 36.2 & 37.5 & 36.7 & 35.0 & 29.6 \\

Qwen2.5-Omni
& 54.4 & 46.8 & 37.3 & 40.0 & 46.4 & 43.4 & 35.3 & 37.1 & 50.7 & 56.4 & 43.4 & 58.9 & 35.4 & 38.7 & 45.9 & 43.9 & 42.2 & 55.1 & 34.1 & 36.3 \\
\rowcolor{lightlight-blue}
\hspace{2pt} + visual input
& 67.1 & 59.1 & 50.4 & 48.0 & 60.9 & 55.9 & 47.5 & 45.7 & 63.7 & 72.0 & 59.2 & 66.1 & 44.7 & 48.8 & 60.8 & 62.6 & 53.1 & 69.4 & 37.5 & 46.0 \\

Gemma 3n E4B IT
& 52.8 & 33.9 & 29.7 & 24.8 & 44.0 & 27.6 & 28.4 & 18.6 & 48.3 & 50.8 & 31.6 & 55.4 & 45.8 & 25.8 & 43.5 & 34.9 & 21.9 & 53.1 & 46.3 & 21.8 \\
\rowcolor{lightlight-blue}
\hspace{2pt} + visual input
& 57.6 & 43.3 & 37.6 & 32.4 & 49.6 & 39.3 & 35.6 & 29.8 & 54.2 & 53.6 & 47.4 & 64.3 & 48.9 & 34.3 & 49.2 & 39.9 & 40.6 & 63.3 & 52.5 & 31.2 \\

Qwen3-Omni
& 65.9 & 61.4 & 46.6 & 46.7 & 59.9 & 57.2 & 41.2 & 44.3 & 63.2 & 70.3 & 51.3 & 71.4 & 47.9 & 45.6 & 59.8 & 61.9 & 45.3 & 71.4 & 41.5 & 41.5 \\
\rowcolor{lightlight-blue}
\hspace{2pt} + visual input
& 67.4 & 59.1 & 54.7 & 49.0 & 61.5 & 55.2 & 50.5 & 46.8 & 64.3 & 72.3 & 56.6 & 67.9 & 44.7 & 51.2 & 61.2 & 63.1 & 51.6 & 71.4 & 37.5 & 47.6 \\

\hline
Gemini 2.5 Flash \dag
& 68.1 & 58.5 & 47.5 & 55.8 & 61.9 & 53.8 & 42.2 & 54.2 & 71.3 & 64.2 & 55.3 & 65.5 & 47.9 & 51.6 & 67.8 & 52.1 & 51.6 & 64.6 & 43.9 & 48.2 \\
\rowcolor{lightlight-blue}
\hspace{2pt} + visual input
& 72.4 & 67.8 & 55.7 & 58.1 & 67.3 & 65.5 & 53.0 & 55.7 & 75.8 & 71.8 & 68.0 & 67.9 & 42.6 & 56.1 & 72.7 & 63.0 & 66.7 & 71.4 & 37.5 & 53.4 \\
\rowcolor{lightlight-gray}
\hspace{2pt} \textit{- audio (text-only)}
& 38.3 & 30.4 & 28.8 & 27.6 & 21.6 & 17.9 & 17.6 & 21.6 & 31.1 & 44.0 & 35.5 & 32.1 & 37.5 & 25.8 & 18.9 & 21.2 & 23.4 & 22.4 & 26.8 & 16.6 \\

Gemini 3 Flash \dag
& 78.7 & 74.3 & 57.6 & 55.2 & 74.3 & 69.7 & 57.8 & 53.6 & 83.0 & 78.6 & 77.6 & 67.9 & 43.8 & 57.1 & 80.4 & 71.4 & 73.4 & 65.3 & 39.0 & 56.5 \\
\rowcolor{lightlight-blue}
\hspace{2pt} + visual input
& 82.7 & 78.4 & 74.6 & 69.5 & 79.4 & 74.5 & 74.5 & 67.0 & 84.9 & 82.7 & 85.5 & \textbf{78.6} & 56.2 & 71.9 & 83.9 & 76.7 & 82.8 & \textbf{75.5} & 51.2 & 70.5 \\

Gemini 3 Pro \dag
& 83.2 & \textbf{82.5} & 66.9 & 65.7 & 80.5 & 80.0 & 65.7 & 62.9 & \textbf{88.3} & 82.0 & 82.9 & 67.9 & \textbf{59.6} & 68.2 & \textbf{87.5} & 76.3 & 82.8 & 65.3 & 55.0 & 66.3 \\
\rowcolor{lightlight-blue}
\hspace{2pt} + visual input
& \textbf{84.7} & 81.3 & \textbf{76.1} & \textbf{72.4} & \textbf{82.6} & \textbf{79.3} & \textbf{76.2} & \textbf{71.1} & 86.6 & \textbf{84.6} & \textbf{90.8} & 71.4 & 58.3 & \textbf{75.0} & 85.8 & \textbf{79.4} & \textbf{92.2} & 71.4 & \textbf{56.1} & \textbf{74.5} \\

\bottomrule
\end{tabular}
}
\caption{Model performance across sound types and languages for \texttt{meme-full} and \texttt{meme-main}.}
\label{tab:main_ST_lang}
\vspace{-0.7cm}
\end{table*}

\vspace{2pt}
\noindent
\textcolor{black!60}{$\ulcorner$The last two types cover the human conventions and external knowledge that anchor the clip in its communicative and cultural setting.$\lrcorner$}
\vspace{2pt}

\noindent
\textbf{\textcolor{medium-blue}{U}sage \& Application} focuses on how people actually use the clip as a meme: the situations it fits, the message it conveys, or the reaction it arises. They test whether the model understands the application of it. \underline{Example}: \textit{When do people use this meme?} %and \textit{Which of the following comments is the most appropriate comment for this video?}

\noindent
\textbf{\textcolor{medium-purple}{W}orld Knowledge} requires information beyond the clip, such as who the speaker is, the media it comes from, or what event or reference it points to. Correct answers rely on cultural familiarity or factual background. \underline{Example}: \textit{Who performs the original version of this track?} %and \textit{Which line is a quote from the original work this song comes from?}

While we define seven question types for clarity of analysis, their boundaries are not absolute. Human interpretation is naturally subjective: for instance, an experienced musician may infer the genre or composer from a first listen (contextual inference), whereas an ordinary listener might need a program book to learn it (world knowledge). These categories should therefore be viewed as guiding notions rather than strict divisions.

\vspace{-0.2cm}
\section{Main Results}
\label{sec:results}

We evaluate 19 state-of-the-art MLLMs: ten audio-only \cite{tang2024salmonn, qwen2audio, audio-reasoner, kimi-audio, audio_flamingo_3, step-audio2, midashenglm, mimo-audio, music_flamingo} including GPT-4o Audio \cite{gpt4o} and nine audio-visual \cite{baichuan-omni, minicpm, phi4mm, qwen2.5omni, gemma_3n, qwen3omni} including Gemini 2.5 Flash \cite{gemini2.5}, Gemini 3 Flash \cite{gemini3-flash}, and Gemini 3 Pro \cite{gemini3-pro}. The exact checkpoints and code repositories if public are documented in Table \ref{tab:model_details}. All models receive the same prompt: question with randomly shuffled options starting from letter \textit{A}. Audio is converted to monaural 16kHz, and video is converted to 360p resolution (aspect ratio preserved) at one frame per second. For GPT and Gemini, we create temporary copies of the audio and video files (\textit{tmp.wav/mp4}) and upload them as bytes, preventing the models from cheating with file names. AVMeme Exam is organized that each Q\&A belongs to a specific question type, and each clip is associated with both a sound type and a language. Leveraging this structure, we report results by question type in Table \ref{tab:main_QT} and by sound category and language in Table \ref{tab:main_ST_lang}, enabling a multi-axis analysis of model behavior and relative strengths and weaknesses. We summarize the key findings below. Supplementary results are shown in Appendix \ref{app:results}.

\vspace{-0.25cm} \noindent\rule{\columnwidth}{0.3pt}\vspace{-0.25cm}

\noindent
\textbf{i. Overall performance.} For both test splits, more recent models (lower in the tables) achieve higher performance. Closed-source commercial models ($\dagger$) significantly outperform open-source models. Gemini 3 Pro is overall the best model in our benchmark, which achieves an average accuracy of 76.6 (audio-only) and 80.0 (audio--visual) on \texttt{meme-main}. The strongest open-source model, Qwen3-Omni, reaches only 55.4 and 57.4, respectively, and still lags behind Gemini 2.5 Flash, despite being considerably stronger than Gemma 3n, Qwen2.5-Omni, and phi-4-Multimodal. For the audio-only, GPT-4o Audio remains the strongest, achieving an average accuracy of 50.1 on \texttt{meme-main}. Across all model families, audio-visual models consistently outperform their audio-only counterparts, confirming the benefit of visual signals for interpreting complex Internet clips, where video provides cues about object motion, scene setting, speaker identity, and interactions that are difficult to infer from audio alone. Finally, accuracies on \texttt{meme-main} are often 5--10\% lower than on \texttt{meme-full} across all models and categories, showing that removing easy questions yields a more challenging evaluation that better tests genuine multimodal understanding.

\vspace{-0.25cm} \noindent\rule{\columnwidth}{0.3pt}\vspace{-0.25cm}

\noindent
\textbf{ii. Content versus context and culture understanding}.
Across nearly all models, Language Analysis (L) is the easiest category, with leading systems such as Gemini 2.5 Flash, Qwen3-Omni, and GPT-4o~Audio reaching 76--83\% on \texttt{meme-main}. Audio Analysis (A) is noticeably harder: even some strongest models are around or below 60\%, reflecting the challenge of interpreting prosody, musical style, and other non-linguistic acoustic cues not captured in speech transcription. Performance drops further for higher-level question types such as Contextual Inference (C), Humor \& Popularity (H), Usage \& Application (U), and World Knowledge (W). Among these, W is the most difficult for most models, typically falling in the 20--55\% range, while U also remains low across both audio-only and audio-visual models. Even top audio-visual systems show 15--30\% declines when moving from L to these deeper understanding tasks. Interestingly, Gemini 3 Flash and Pro show significantly higher accuracy on World Knowledge than all other models including 2.5 Flash, with 3 Pro exceeding Qwen3-Omni by over 30\% on W. This advantage may be related to broader exposure to the multimedia world, e.g. YouTube videos and associated metadata, during training. Nevertheless, Context and Usage questions still remain the hardest, with performance declining sharply from L (above 90\%). Taken together, a finding is consistent across all models: while MLLMs excel in surface semantic parsing, contextual, cultural, and pragmatic understanding still remains challenging.

\vspace{-0.25cm} \noindent\rule{\columnwidth}{0.3pt}\vspace{-0.25cm}

\noindent
\textbf{iii. Challenge of textless sounds}. Table \ref{tab:main_ST_lang} highlights another consistent bottleneck. Models perform best on speech (Sp), good on songs (So), worse on music (Mu), and worst on sound effects (Sfx). The last two types typically have no spoken words. Even the strongest audio LLMs, such as GPT-4o and Music Flamingo, reach only 35 to 45\% on music and sound effects, compared with around 60 to 65\% on speech and songs. Certain models exhibit specialized strengths; for example, Kimi-Audio and Audio Flamingo 3 achieve higher accuracy than GPT-4 on sound effects or music despite being weaker overall. These patterns suggest that current MLLMs handle textual semantics well, but their ability to interpret music and sound effects without linguistic structure falls largely behind.

\vspace{-0.25cm} \noindent\rule{\columnwidth}{0.3pt}\vspace{-0.25cm}

\noindent
\textbf{iv. Challenge of lesser-known languages}.
Performance also varies significantly across languages. English and Chinese are scored best for nearly all models. Qwen3-Omni is particularly strong in Chinese, outperforming Gemini 2.5 Flash, while the two models perform similarly in English. Japanese, Korean, and Persian are substantially more difficult, with leading models frequently dropping to the 35–55\% range and audio-only LLMs sometimes falling below 30\% . Even Gemini 3 pro only scores 56.1\% in Persian on \texttt{meme-main}. The \textit{nan} columns in Table \ref{tab:main_ST_lang} denote clips without language, mainly music and sound effects, whose performance is comparable to or even lower than Persian depending on the model.
Adding visual input improves accuracy only marginally for these lesser-known languages and non-verbal sounds.

\vspace{-0.25cm} \noindent\rule{\columnwidth}{0.3pt}\vspace{-0.25cm}

\noindent
\textbf{v. MLLMs vs humans in meme comprehension}.
To provide models a human reference, we recruit 20 participants, 10 native English speakers and 10 native Chinese speakers aged 18–35, all self-reported frequent Internet users. Each participant answers only clips in their native language and culture, or clips without language, without web search or collaboration. This narrows to 750 memes in total and 37.5 memes per person. Notice that this setting \textbf{\textit{disadvantages}} humans, as unlike LLMs trained on web data, no individual has seen all videos. Yet the strong performance of individual humans despite this disadvantage \textbf{\textit{strengthens}} our takeaway. Before showing them Q\&A, we first ask participants whether they have seen and are familiar with each meme video, and report results grouped by human familiarity level in Figure~\ref{fig:human_vs_llm} (More details in Appendix \ref{app:human}). As expected, humans perform much better on their familiar memes, outperforming \textit{\textbf{all}} open-source MLLMs and comparable to Gemini 2.5 Flash. Qwen3-Omni, the strongest open-source model, only slightly outperforms a human individual when we count the memes they have not seen. But even on unseen memes, humans still outperform the remaining open-source MLLMs, showing the continued advantage of human cultural grounding in meme understanding.

\vspace{-0.25cm} \noindent\rule{\columnwidth}{0.3pt}\vspace{-0.25cm}

\noindent
\textbf{vi. Does thinking help?} We compare Gemini 2.5/3 Flash with minimal thinking and Gemini~3.0~Pro with low and high thinking levels (Table \ref{tab:gemini} in the Appendix). At an aggregate level, thinking appears beneficial: high thinking improves AVG accuracy by 5.4\% in the audio-only setting and 1.7\% in the audio-visual setting on \texttt{meme-main}. However, category-level trends are highly uneven, with gains concentrated in explicit recognition tasks. In audio-only mode, Audio improves from 57.8 to 71.1 and Language from 83.5 to 94.2, while World Knowledge also increases from 68.8 to 77.9, with more \textit{``Aha, I know the correct video!''} in longer thinking. However, longer thinking offers little or even adverse benefit for pragmatic and cultural tasks. In the audio-visual setting, accuracy stagnates for Context (80.4~$\rightarrow$~80.4) and Humor (75.3~$\rightarrow$~75.3), while Usage drops (72.2~$\rightarrow$~71.0). In audio-only mode, Emotion (81.0~$\rightarrow$~77.6) and Humor (71.8~$\rightarrow$~70.6) also decline. Appendix~\ref{app:think_cases} shows examples of both effective and misguided thinking.

\vspace{-0.25cm} \noindent\rule{\columnwidth}{0.3pt}\vspace{-0.25cm}

% \begin{figure}
%     \centering
%     \includegraphics[width=\linewidth]{figure/familiarity.pdf}
%     \caption{Models vs. human individuals, grouped by humans are faimilar/unfamilar/unseen the meme.}
%     \label{fig:human_vs_llm}
%     \vspace{-0.5cm}
% \end{figure}

\noindent
\textbf{vii. Ablations}. In Table \ref{tab:ablation}, we ablate the evaluation setting to show how easily multimodal evaluation can drift away from genuine understanding. First, when we provide the name of the meme (written by the contributor, which often contains the song title, character name, or the meme itself), model accuracy increases significantly ($\approx$10\%) because many solutions become recallable from the LLM training text, bypassing audio/video understanding. Second, adding the soft hint \textit{This is a meme} before the question slightly boosts accuracy (1--2\%) for most models, as they begin to treat the clip like a familiar Internet template and to think more (especially in ironic ways). While not cheating, we avoid this prefix so that models do not mistake ordinary videos as memes. Finally, allowing visual-cheat conditions, i.e., keeping videos that disclose key information, causes substantial accuracy inflation, 40\% or more on the visual-cheatable subset (numbers in parentheses), because on-screen text, titles, or visual objects directly reveal the answer. These observations justify our strict removal of both text and visual shortcuts for fair multimodal evaluation.

\vspace{-0.25cm} \noindent\rule{\columnwidth}{0.3pt}\vspace{-0.25cm}

\noindent
\textbf{Discussion}. Taken together, our results and analyses reveal a consistent pattern: MLLMs perform reliably on content-level questions, particularly when rich linguistic cues are present. Their performance drops sharply when understanding requires contextual thinking or cultural grounding, and on textless audio or less common languages. Even the strongest models like Gemini 3 Pro exhibit the largest errors in these categories, and longer thinking often does not help. This underscores that pragmatic and cultural comprehension remain open challenges for multimodal intelligence.

\vspace{-0.25cm} \noindent\rule{\columnwidth}{0.3pt}\vspace{-0.25cm}

\noindent
\textbf{Lesson}. Beyond increasing exposure to Internet audio and video, our findings suggest that multimodal intelligence should be more firmly grounded in the human world. Understanding audio–visual signals is not merely about perceiving isolated sounds or images, but about connecting them to how people interpret, feel about, and use them within context and culture. This points to the importance of richer, human-centered annotations that capture emotion, usage, and cultural meaning, along with training objectives that prioritize alignment with human interpretation over surface-level recognition.

\begin{table}[t]
\centering
\resizebox{0.7\linewidth}{!}{
\setlength{\tabcolsep}{2pt}
\begin{tabular}{
p{0.24\linewidth}|
p{0.24\linewidth}  % This is a meme
p{0.24\linewidth}  % Name Given
p{0.34\linewidth}  % Visual-Cheat (all) -- doubled width
|p{0.34\linewidth}  % Default
}
\toprule
\multirow{2}{*}{Model}
& \multicolumn{4}{c}{Evaluation Settings} \\
\cmidrule(lr){2-5}
& \textit{``This is a meme."}+Q
& Meme Name+Q
& Keep videos for \texttt{visual\_cheat}
& Default (minimal hint or cheat)\\
\midrule

Qwen2-Audio
& 35.4 & 44.6 & \textit{n.a.} & 34.4 \\

Audio Flamingo 3
& 41.5 & 50.0 & \textit{n.a.}  & 41.7 \\

Music Flamingo 
& 42.9 & 51.3 & \textit{n.a.}  &  42.8 \\

\midrule

Baichuan-Omni
& 34.1 & 46.1 & \textit{n.a.}  & 33.1 \\
\rowcolor{lightlight-blue}
\hspace{2pt}+ visual input 
& 41.3 & 49.4 & 45.3 (80.3) & 39.6 (40.0) \\
Phi-4-Multimodal
& 34.9 & 42.5 & \textit{n.a.}  & 35.6 \\
\rowcolor{lightlight-blue}
\hspace{2pt}+ visual input 
& 42.4 & 49.5 & 43.6 (74.4) & 38.3 (35.0) \\

Gemma 3n E4B IT
& 38.0 & 53.6 & \textit{n.a.}  & 36.4 \\
\rowcolor{lightlight-blue}
\hspace{2pt}+ visual input 
& 44.3 & 57.1 & 50.4 (84.6) & 43.9 (38.5) \\

Qwen3-Omni
& 53.6 & 67.1 & \textit{n.a.} & 55.4 \\
\rowcolor{lightlight-blue}
\hspace{2pt}+ visual input 
& 58.7 & 67.8 & 61.7 (81.2) & 57.4 (49.6) \\

\bottomrule
\end{tabular}}
\caption{The effect of text or visual \textit{hint} and \textit{cheat} in multimodal evaluation. The right-most column is our default setting with minimal answer shortcuts. Results from \texttt{meme-main}.}
\label{tab:ablation}
\end{table}

\section{Conclusion}

We presented AVMeme Exam, a multimodal, multilingual, and multicultural benchmark that evaluates whether MLLMs can understand not only what is heard or seen, but also how meaning is constructed through context, emotion, usage, and shared cultural knowledge. Our analysis reveals consistent limitations: models perform substantially worse on textless audio, struggle with contextual and cultural reasoning, and often fail to align with human interpretations. These findings highlight a key gap between current multimodal capabilities and the human expectations of intent-aware, culturally grounded understanding. In future work, we plan to expand AVMeme Exam to broader cultures, linguistic communities, and media ecosystems, creating a more globally representative testbed for multimodal understanding. We also aim to develop methods that could address the challenges revealed by our benchmark, advancing progress toward human-aligned multimodal intelligence.

\textbf{Limitations.} While our AVMeme Exam spans diverse languages and media, several limitations remain. First, cultural coverage is inevitably shaped by our contributors, who are highly-educated researchers aged 22--35 and whose perspectives do not fully represent the global, intergenerational landscape of meme culture. Second, meme meaning shifts over time. Our annotations capture contemporary (by the end of 2025) interpretations and cannot anticipate future cultural drift (which is impossible). Third, most current MLLMs impose strict audio and video length limits, requiring clips to be truncated to 30 seconds, which may omit context essential for real-world understanding. Fourth, our evaluation focuses on controlled multiple-choice Q\&A on single clips, whereas real-world applications often require multi-turn dialogue, personalization, and open-ended scenarios. Fifth, meme interpretation is inherently subjective. Although viewers may converge on a majority understanding, alternative readings are equally valid and cannot be fully reflected in a standardized benchmark. Taken together, these considerations suggest that AVMeme Exam should be viewed as a reference benchmark, diagnostic and comparative, rather than an absolute ground truth for human multimedia culture. Nevertheless, the lesson from AVMeme Exam, that current multimodal AIs remain weaker at audio–visual understanding than text and at contextual \& cultural thinking than surface content, still holds.

\section*{Ethical Considerations}

AVMeme Exam is designed as a research benchmark to study multimodal understanding of audio–visual memes, with careful attention to ethical considerations in data collection, annotation, and verification. All audio–visual clips are human-curated by contributors who personally recognize and use these memes, rather than being automatically scraped, and are drawn exclusively from publicly available online videos, with no private, paywalled, or confidential content included. To mitigate potential harm, we prohibit political materials and explicit depictions of sexual, violent, hateful, criminal, or drug-related content, and clearly annotate clips with implicit or contextual sensitivity using sensitivity tags to enable filtering and informed use. The benchmark is intended for research and evaluation, not for safety-critical or high-stakes deployment, and any biases or cultural limitations it reveals should be understood as limitations of current models and of the dataset itself, motivating further work toward more human-aligned, culturally grounded multimodal systems.

Human evaluation was conducted under an Institutional Review Board (IRB) protocol. All participants completed the evaluation online in approximately 30 minutes and were compensated \$15.

\section*{Acknowledgement}

We acknowledge the use of Nano Banana Pro to draw meme figures for illustrative purposes. LLMs were also used in the text-cheat detection procedure described in the paper, and for syntax and grammar fixes. All final data verification and paper writing decisions were made by the human authors.

This work is funded by the National Institutes of Health (NIH-NIDCD) and a grant from Marie-Josee and Henry R. Kravis.

\bibliographystyle{colm2026_conference}
\bibliography{colm2026_conference}
\newpage
\null

\appendix

\section{Collection \& Verification Procedures}
\label{app:collection}

AVMeme Exam was curated by 27 audio and NLP researchers affiliated with universities and companies in the US and the UK, aged between 22 and 35, the majority of whom are current PhD students. To coordinate collection, all raw entries are first recorded in a shared Google Sheet, organized into separate sheets by language and sound category, which helps prevent duplicate submissions across different contributors. The complete raw data collection process took approximately three months.

All contributors participated voluntarily and with informed consent. They are listed as co-authors of the paper. To mitigate potential ethical risks in data curation, contributors were instructed to rely only on publicly available content, avoid political or religious material, and exclude explicit sexual, violent, or hateful content. No personal data of contributors was collected or released.

All contributors were given and agreed to follow the collection and annotation guidelines:
\begin{itemize}[noitemsep]
    \item \textbf{Meme selection:} Memes are chosen based on broad public recognition and sourced from original or highly circulated online videos, preferably from YouTube. A view count of at least one million is a soft requirement.
    \item \textbf{Clip extraction:} Each meme is trimmed to a short segment (1--30 seconds). At most two distinct memes may be extracted from the same source video.
    \item \textbf{Audio-centric requirement:} Memes must be identifiable primarily from audio alone. Clips whose meaning relies heavily on visual information are excluded.
    \item \textbf{Multilingual and cross-cultural coverage:} Memes may span multiple languages and cultures, and are explicitly labeled while retaining their original sound category.
    \item \textbf{Standardized metadata:} Each entry includes structured annotations such as source URLs, temporal boundaries, original appearance date, language, sound category, transcription (when applicable), summary, emotion, sensitivity, and typical usage.
    \item \textbf{Content safeguards:} Political and religious content, explicit sexual material, and graphic violence are excluded. Mild or implicit sensitive content is allowed only with explicit sensitivity labels.
    \item \textbf{Question design:} Each meme is paired with a unique multiple-choice question targeting different levels of understanding. Contributors are encouraged to design challenging and creative questions, and are not required to assign questions to any predefined categories. Question types are labeled later during verification by independent annotators.
    \item \textbf{Leakage prevention:} Questions and answer choices must not disclose identifying information (e.g., titles, names, slogans, or sources) to prevent textual shortcuts and ensure genuine audio-visual evaluation.
\end{itemize}

\begin{table*}[t]
\centering
\small
\setlength{\tabcolsep}{1pt}
\resizebox{\linewidth}{!}{%
\begin{tabular}{c|ccc}
\toprule
\textbf{Model} & \textbf{Checkpoint / Version} & \textbf{Size} & \textbf{Code Repository} \\
\midrule
SALMONN & SALMONN-13B & 13B & https://huggingface.co/tsinghua-ee/SALMONN \\
Qwen2-Audio & Qwen2-Audio-7B & 7B & https://huggingface.co/Qwen/Qwen2-Audio-7B \\
Audio-Reasoner & Audio-Reasoner-7B & 7B & https://huggingface.co/zhifeixie/Audio-Reasoner \\
Kimi-Audio & Kimi-Audio-7B-Instruct & 7B & https://huggingface.co/moonshotai/Kimi-Audio-7B-Instruct \\
Audio Flamingo 3 & audio-flamingo-3-hf & 7B & https://huggingface.co/nvidia/audio-flamingo-3 \\
Step-Audio-2 & Step-Audio-2-mini & 8B & https://huggingface.co/stepfun-ai/Step-Audio-2-mini \\
MiDashengLM & midashenglm-7b-0804-fp32 & 7B & https://huggingface.co/mispeech/midashenglm-7b-0804-fp32 \\
MiMo-Audio & MiMo-Audio-7B-Instruct & 7B & https://huggingface.co/XiaomiMiMo/MiMo-Audio-7B-Instruct \\
Music Flamingo & music-flamingo-hf & 7B & https://huggingface.co/nvidia/music-flamingo-hf \\
GPT-4o Audio & gpt-4o-audio-preview-2025-06-03 & - & - \\
Baichuan-Omni & Baichuan-Omni-1.5 & 7B & https://github.com/baichuan-inc/Baichuan-Omni-1.5 \\
MiniCPM & MiniCPM-o-2\_6 & 8B & https://huggingface.co/openbmb/MiniCPM-o-2\_6 \\
phi-4-Multimodal & Phi-4-multimodal-instruct & 5.6B & https://huggingface.co/microsoft/Phi-4-multimodal-instruct \\
Qwen2.5-Omni & Qwen2.5-Omni-7B & 7B & https://huggingface.co/Qwen/Qwen2.5-Omni-7B \\
Gemma 3n E4B IT & gemma-3n-E4B-it & 8B & https://huggingface.co/google/gemma-3n-E4B-it \\
Qwen3-Omni & Qwen3-Omni-30B-A3B-Instruct & 30B & https://huggingface.co/Qwen/Qwen3-Omni-30B-A3B-Instruct \\
Gemini 2.5 Flash & gemini-2.5-flash & - & - \\
Gemini 3 Flash & gemini-3-flash-preview & - & - \\
Gemini 3 Pro & gemini-3-pro-preview & - & - \\
\bottomrule
\end{tabular}%
}
\caption{Details about model checkpoints and repositories. We follow the provided inference code with default configurations for each model.}
\label{tab:model_details}
\end{table*}

\begin{table*}[!h]
\centering

%%%%%%%%%%%%%%%%%%%%%%%%%%%%
% Part A: Question Types
%%%%%%%%%%%%%%%%%%%%%%%%%%%%
\resizebox{\textwidth}{!}{
\setlength{\tabcolsep}{4pt}
\begin{tabular}{
p{0.26\textwidth}|
*{8}{>{\centering\arraybackslash}p{0.052\textwidth}}|
*{8}{>{\centering\arraybackslash}p{0.052\textwidth}}
}
\toprule
\multirow{2}{*}{Gemini (\texttt{thinking\_level})}
& \multicolumn{8}{c|}{\texttt{meme-full}}
& \multicolumn{8}{c}{\texttt{meme-main}} \\
\cmidrule(lr){2-9} \cmidrule(lr){10-17}
& A & L & C & E & H & U & W & AVG
& A & L & C & E & H & U & W & AVG \\
\midrule

2.5 Flash (\texttt{minimal})
& 51.8 & 85.4 & 61.6 & 65.8 & 68.2 & 60.0 & 53.3 & 62.9
& 42.2 & 83.3 & 55.7 & 58.6 & 53.6 & 56.2 & 51.0 & 57.2 \\
\rowcolor{lightlight-blue}
\hspace{2pt} + visual input
& 56.4 & 82.3 & 69.2 & 78.1 & 72.2 & 60.6 & 63.6 & 68.3
& 47.7 & 79.6 & 66.1 & 72.4 & 60.0 & 57.1 & 62.1 & 63.9 \\

3 Flash (\texttt{minimal})
& 64.3 & 86.3 & 75.4 & 75.3 & 77.4 & 67.7 & 67.4 & 73.2
& 55.6 & 84.5 & 71.4 & 72.4 & 67.1 & 64.2 & 66.3 & 69.1 \\
\rowcolor{lightlight-blue}
\hspace{2pt} + visual input
& 64.3 & 88.7 & 80.6 & 82.2 & 84.2 & 74.4 & 78.9 & 79.7
& 55.6 & 87.4 & 78.8 & 79.3 & 76.5 & 70.4 & 77.9 & 76.6 \\

3 Pro (\texttt{low})
& 66.1 & 85.5 & 74.1 & 84.9 & 79.7 & 70.3 & 70.0 & 74.9
& 57.8 & 83.5 & 71.4 & 81.0 & 71.8 & 66.0 & 68.8 & 71.2 \\
\rowcolor{lightlight-blue}
\hspace{2pt} + visual input
& 76.8 & 89.5 & 82.4 & 86.3 & 82.7 & 74.9 & 77.9 & 80.9
& 71.1 & 89.3 & 80.4 & 82.8 & 75.3 & 72.2 & 77.3 & 78.3 \\

3 Pro (\texttt{high})
& 76.8 & 94.4 & 77.8 & 80.8 & 78.2 & 73.7 & 78.9 & 79.5
& 71.1 & 94.2 & 74.7 & 77.6 & 70.6 & 70.2 & 77.9 & 76.6 \\
\rowcolor{lightlight-blue}
\hspace{2pt} + visual input
& 80.4 & 91.1 & 81.1 & 86.3 & 83.5 & 72.8 & 83.6 & 81.9
& 75.6 & 90.3 & 80.4 & 84.5 & 75.3 & 71.0 & 83.1 & 80.0 \\

\bottomrule
\end{tabular}
}

\vspace{6pt}

%%%%%%%%%%%%%%%%%%%%%%%%%%%%
% Part B: Sound Types & Language
%%%%%%%%%%%%%%%%%%%%%%%%%%%%
\resizebox{\textwidth}{!}{
\setlength{\tabcolsep}{4pt}
\begin{tabular}{
p{0.26\textwidth}|
*{4}{>{\centering\arraybackslash}p{0.036\textwidth}}|
*{4}{>{\centering\arraybackslash}p{0.036\textwidth}}|
*{6}{>{\centering\arraybackslash}p{0.036\textwidth}}|
*{6}{>{\centering\arraybackslash}p{0.036\textwidth}}
}
\toprule
\multirow{2}{*}{Gemini (\texttt{thinking\_level})}
& \multicolumn{4}{c|}{\texttt{meme-full}: ST}
& \multicolumn{4}{c|}{\texttt{meme-main}: ST}
& \multicolumn{6}{c|}{\texttt{meme-full}: Lang}
& \multicolumn{6}{c}{\texttt{meme-main}: Lang} \\
\cmidrule(lr){2-5}\cmidrule(lr){6-9}
\cmidrule(lr){10-15}\cmidrule(lr){16-21}
& Sp & So & Mu & Sfx
& Sp & So & Mu & Sfx
& en & zh & ja & ko & fa & nan
& en & zh & ja & ko & fa & nan \\
\midrule

2.5 Flash (\texttt{minimal})
& 68.1 & 58.5 & 47.5 & 55.8
& 61.9 & 53.8 & 42.2 & 54.2
& 71.3 & 64.2 & 55.3 & 65.5 & 47.9 & 51.6
& 67.8 & 52.1 & 51.6 & 64.6 & 43.9 & 48.2 \\
\rowcolor{lightlight-blue}
\hspace{2pt} + visual input
& 72.4 & 67.8 & 55.7 & 58.1
& 67.3 & 65.5 & 53.0 & 55.7
& 75.8 & 71.8 & 68.0 & 67.9 & 42.6 & 56.1
& 72.7 & 63.0 & 66.7 & 71.4 & 37.5 & 53.4 \\

3 Flash (\texttt{minimal})
& 78.7 & 74.3 & 57.6 & 55.2
& 74.3 & 69.7 & 57.8 & 53.6
& 83.0 & 78.6 & 77.6 & 67.9 & 43.8 & 57.1
& 80.4 & 71.4 & 73.4 & 65.3 & 39.0 & 56.5 \\
\rowcolor{lightlight-blue}
\hspace{2pt} + visual input
& 82.7 & 78.4 & 74.6 & 69.5
& 79.4 & 74.5 & 74.5 & 67.0
& 84.9 & 82.7 & 85.5 & 78.6 & 56.2 & 71.9
& 83.9 & 76.7 & 82.8 & 75.5 & 51.2 & 70.5 \\

3 Pro (\texttt{low})
& 79.2 & 73.1 & 64.4 & 63.8
& 75.5 & 70.3 & 60.8 & 60.8
& 83.3 & 79.3 & 68.4 & 67.9 & 62.5 & 63.1
& 82.4 & 72.0 & 67.2 & 67.3 & 58.5 & 59.6 \\
\rowcolor{lightlight-blue}
\hspace{2pt} + visual input
& 83.5 & 79.5 & 78.6 & 69.5
& 80.6 & 79.3 & 76.2 & 67.0
& 86.9 & 85.0 & 82.9 & 71.4 & 52.1 & 73.6
& 86.6 & 80.4 & 84.4 & 69.4 & 46.3 & 70.8 \\

3 Pro (\texttt{high})
& 83.2 & 82.5 & 66.9 & 65.7
& 80.5 & 80.0 & 65.7 & 62.9
& 88.3 & 82.0 & 82.9 & 67.9 & 59.6 & 68.2
& 87.5 & 76.3 & 82.8 & 65.3 & 55.0 & 66.3 \\
\rowcolor{lightlight-blue}
\hspace{2pt} + visual input
& 84.7 & 81.3 & 76.1 & 72.4
& 82.6 & 79.3 & 76.2 & 71.1
& 86.6 & 84.6 & 90.8 & 71.4 & 58.3 & 75.0
& 85.8 & 79.4 & 92.2 & 71.4 & 56.1 & 74.5 \\

\bottomrule
\end{tabular}
}

\caption{Gemini family performance across question types, sound types, and languages.}
\label{tab:gemini}
\end{table*}

For leakage prevention, we label Q\&As that can be correctly guessed by all of Gemini 2.5 Flash, Grok 4, and GPT-5.1 in text-only settings, with prompt “\textit{without audio, guess the best option}.”, and return them to the original contributors for revision. This process is repeated for two rounds.

\vspace{-0.25cm} \noindent\rule{\columnwidth}{0.3pt}\vspace{-0.25cm}

AVMeme Exam is verified by nine human verifiers. To reduce bias and confirmation effects, we assign each meme to verifiers who did not author the corresponding entry whenever possible. For less common languages (e.g., Persian and Telugu), where linguistic expertise is limited, original contributors assist in verification to ensure correctness.

The verification process is conducted in a distributed manner through a web-based interface. For each meme, verifiers review all annotated attributes and the multiple-choice Q\&A. For each attribute, verifiers select one of several actions: \textit{Ok}, \textit{Wrong}, or \textit{Wrong, but I can fix}. Identified issues are either corrected directly by the verifier (if just a simple typo or video-question mismatch) or returned to the original contributor for revision.

After validating all metadata fields, verifiers assign the question to one of the predefined question types. Only after completing all checks and assigning a question type can a verifier proceed to the next sample. This structured, attribute-level verification ensures annotation consistency, correctness, and alignment with the intended evaluation goals.

Finally, one human verifier aggregates all verified entries into a single file, manually reviews the merged results to remove duplicate entries, and performs a final pass to ensure consistency in question type assignments across the dataset.

\begin{figure*}
    \centering
    \includegraphics[width=\linewidth]{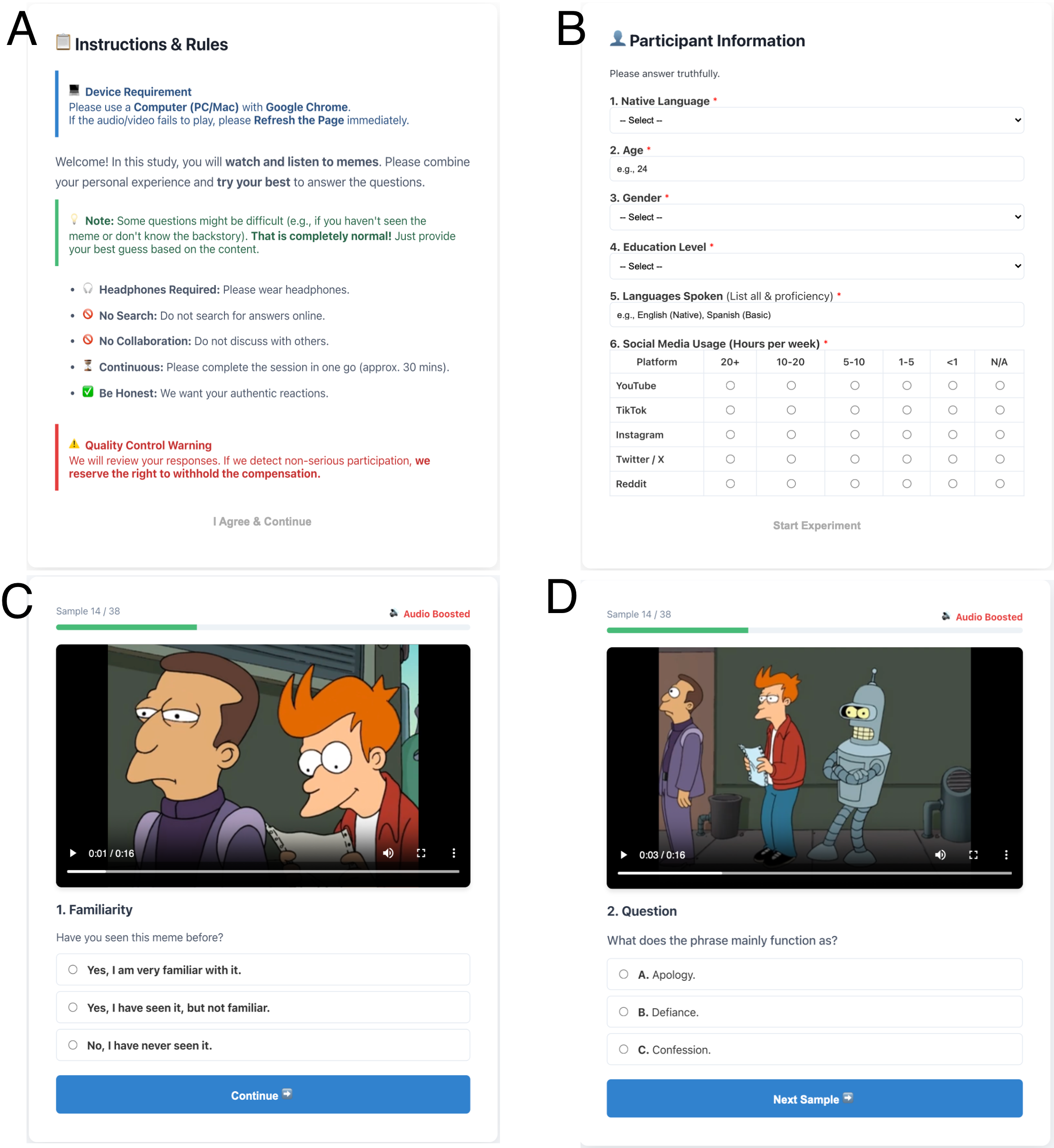}
    \caption{Human evaluation interface. \textbf{A}. Instructions and rules shown before the study, specifying device requirements, headphone use, no-search/no-collaboration constraints, and quality control policies to ensure reliable human judgments. \textbf{B}. Participant information and background survey collected anonymously and linked only via a participant ID, including demographics, language background, and social media usage. \textbf{C}. Familiarity check presented before showing the Q\&A, where participants indicate whether they have previously seen the meme clip. \textbf{D}. Q\&A, in which participants watch and listen to the meme clip and answer a multiple-choice question.}

    \label{fig:human_eval}
\end{figure*}

\section{Human Evaluation Details}
\label{app:human}

We conduct a controlled human evaluation to establish a human reference for the AVMeme Exam, following the University's Institutional Review Board (IRB) protocol. A total of 20 participants were recruited, including 10 native English speakers and 10 native Chinese speakers, all aged 18--35 and self-reported frequent users of online video platforms (e.g., YouTube, Bilibili, TikTok). Participants grew up in the U.S. or China and were residing in the U.S. at the time of the study, due to practical recruitment constraints. All participants completed the study individually, without web search, collaboration, or external assistance.

Each evaluation session lasted approximately 30 minutes, and participants were compensated with a \$15 prepaid card upon completion. Each participant was assigned 37 or 38 audio-visual meme clips, resulting in 750 total human-evaluated samples. To ensure linguistic and cultural appropriateness, participants only evaluated clips in their native language and cultural context, as well as clips without spoken language (e.g., music or sound effects). Screenshots of the session are shown in Figure \ref{fig:human_eval}.

Before the main evaluation, participants completed a background survey collecting demographic information (age range, native language), Internet usage frequency, and self-reported familiarity with online memes and viral media. Each participant was assigned a unique anonymous identifier, and no personally identifiable information was collected. Survey responses were linked only to these anonymous identifiers and were used solely to describe the participant pool and report aggregate evaluation metadata.

For each sample, participants were first shown the video clip alone (or the audio clip for the visual cheating condition, as we enforce for LLMs), without the associated question or answer options. They were asked whether they had previously seen and were familiar with the clip. Only after recording this response were participants presented with the corresponding multiple-choice question and answer options. This two-stage design prevents participants' familiarity judgments from being influenced by question difficulty.

Participants then answered the same multiple-choice questions used for model evaluation, under identical answer formats and constraints. Human responses were recorded and later aggregated by familiarity category for comparison with model performance, as illustrated in Fig.~\ref{fig:human_vs_llm}. This protocol ensures that human evaluation is conducted under controlled, reproducible conditions aligned with the LLM evaluation setting.

\section{Supplementary Results}
\label{app:results}

\begin{figure}[!h]
    \centering
\includegraphics[width=0.55\linewidth]{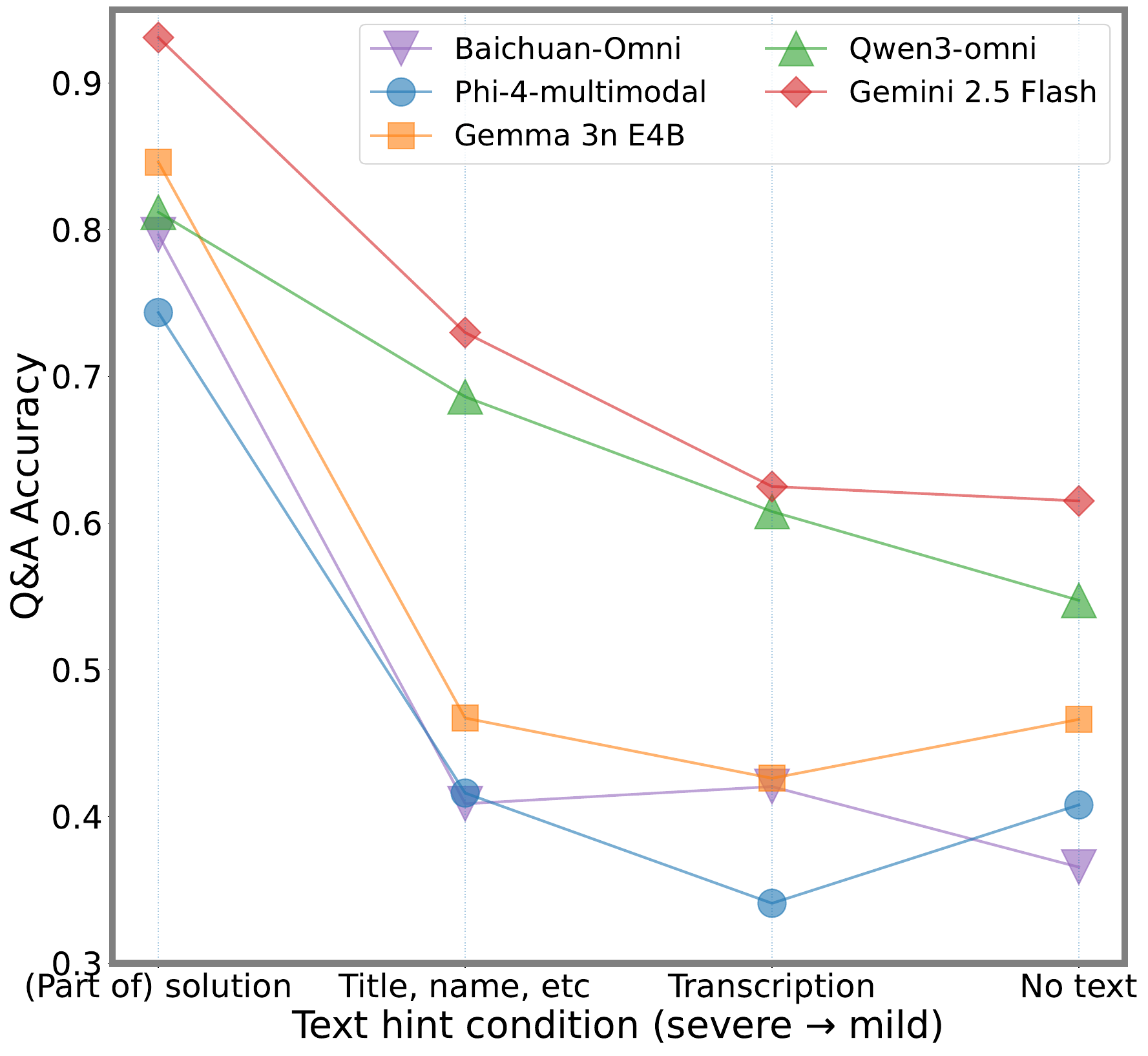}
    \caption{The effect of different levels of text hint from the visual. Results from \texttt{meme-main}.}
    \label{fig:text_hint}
\end{figure}

\begin{figure}[!h]
    \centering
    \includegraphics[width=0.55\linewidth]{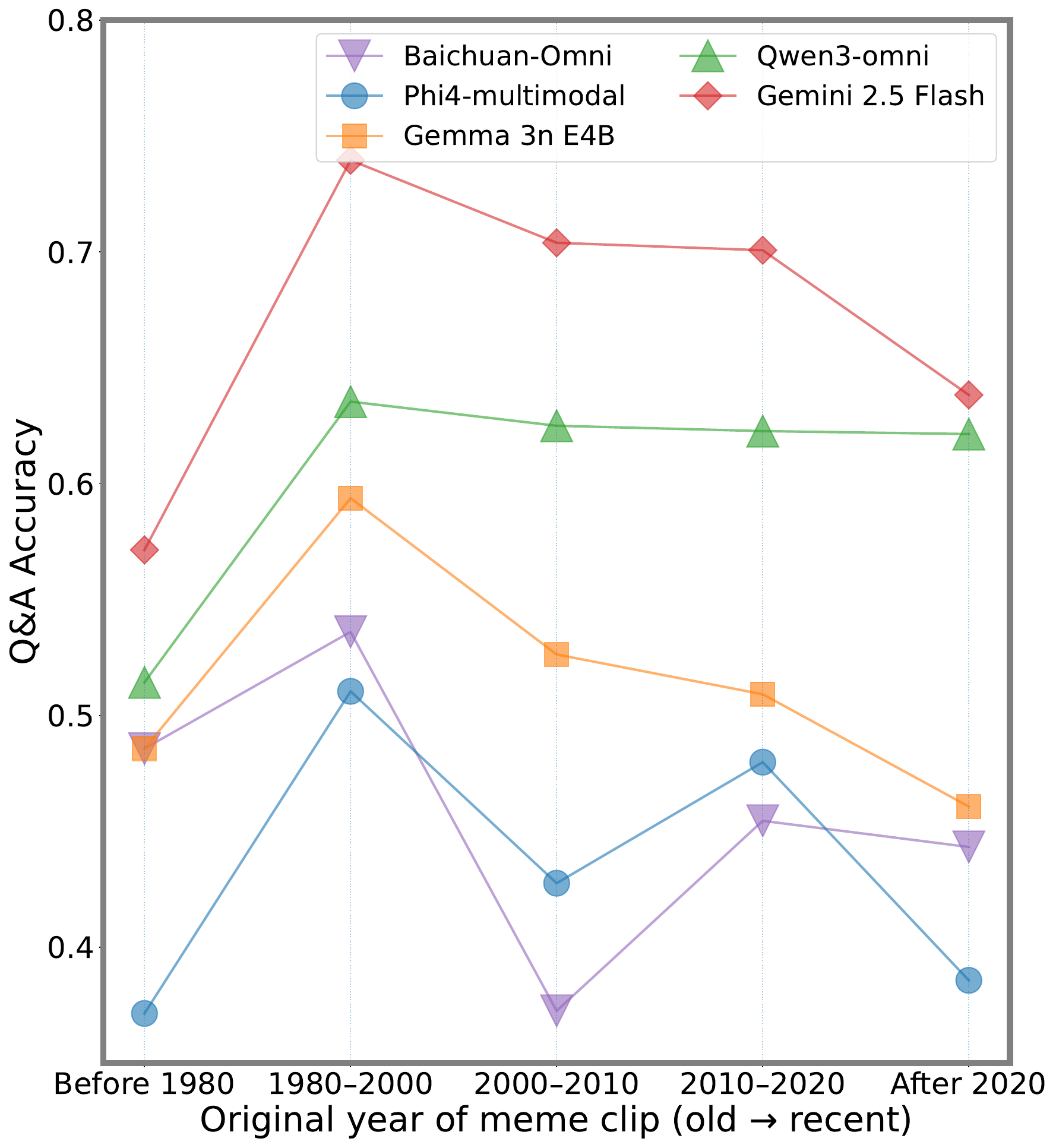}
    \caption{Model performance vs. the original year of the clips. Results from \texttt{meme-main}.}
    \label{fig:meme_year}
    \vspace{-0.5cm}
\end{figure}

\subsection{Study on On-screen Text}
\label{app:onscreen_text}

Real-world videos often contain on-screen text that may partially assist or hack multimodal understanding. To analyze how such visual text affects model performance, we conduct a focused study on clips with different degrees of on-screen textual hints.

We (9 human verifiers) manually label each clip based on the strongest type of text appearing in the visual stream. Specifically, we categorize clips into five groups: (1) \textit{visual contains (part of) solution} (\texttt{visual\_cheat}) (2) \textit{title, name, or other keywords}, which reveal identifiable entities (3) \textit{transcription}, where the spoken content or its translation is displayed as subtitles (4) \textit{no text}. These categories form a spectrum of visual hints, ranging from the most severe to the mildest.

For each category, we aggregate model accuracy across all applicable clips and report the results in Figure~\ref{fig:text_hint}. Accuracy peaks when the visual stream directly reveals the solution and decreases monotonically as text hints weaken from titles and keywords to transcription and ultimately to no text. This pattern is consistent across models, indicating that on-screen text acts as a strong shortcut. For some models, the no-text condition is slightly worse than transcription, as many textless clips involve music or sound effects, which are inherently more challenging.

\subsection{Study on Meme's Year}

We additionally annotate each meme clip with its original year, defined as the upload year of the source YouTube or Bilibili video, or the debut or premiere year for music, songs, and movies. We group clips into coarse temporal spans and aggregate model accuracy within each group, as shown in Figure \ref{fig:meme_year}.

Across models, performance peaks on memes originating between 1980 and 2000, while accuracy drops for both older ($<$1980) and more recent ($>$2020) clips. We hypothesize that this pattern reflects uneven Internet coverage: middle-era memes are well represented and repeatedly circulated online, whereas very old or very recent content is less consistently captured in training data.

\subsection{Correct and Wrong Thinking Example}
\label{app:think_cases}

In addition to the quantitative results of Gemini~3 Flash and Gemini~3 Pro under low and high thinking levels in Table~\ref{tab:gemini}, we present examples below illustrating both effective and misguided reasoning (all in audio-visual mode), in supplement to the discussion in Section \ref{sec:results}.vi.\textit{Does thinking help?}

\noindent
\begin{minipage}[t]{0.48\textwidth}
\begin{memeexample}[colback=gray!2,colframe=gray,
  title={GTA wasted}, left=4pt,right=4pt,top=4pt,bottom=4pt]
\small
\color{black!70}

\noindent
\begin{minipage}[t]{0.58\textwidth}
    \vspace{0pt}
    \textbf{\textcolor{black}{Category:}} sound effect \\
    \textbf{\textcolor{black}{Year:}} 2013 \\
    \textbf{\textcolor{black}{Language:}} N/A \\
    \textbf{\textcolor{black}{Transcription:}} N/A
\end{minipage}%
\hfill
\begin{minipage}[t]{0.32\textwidth}
    \vspace{0pt}
    \raggedleft
    \includegraphics[width=\textwidth, keepaspectratio]{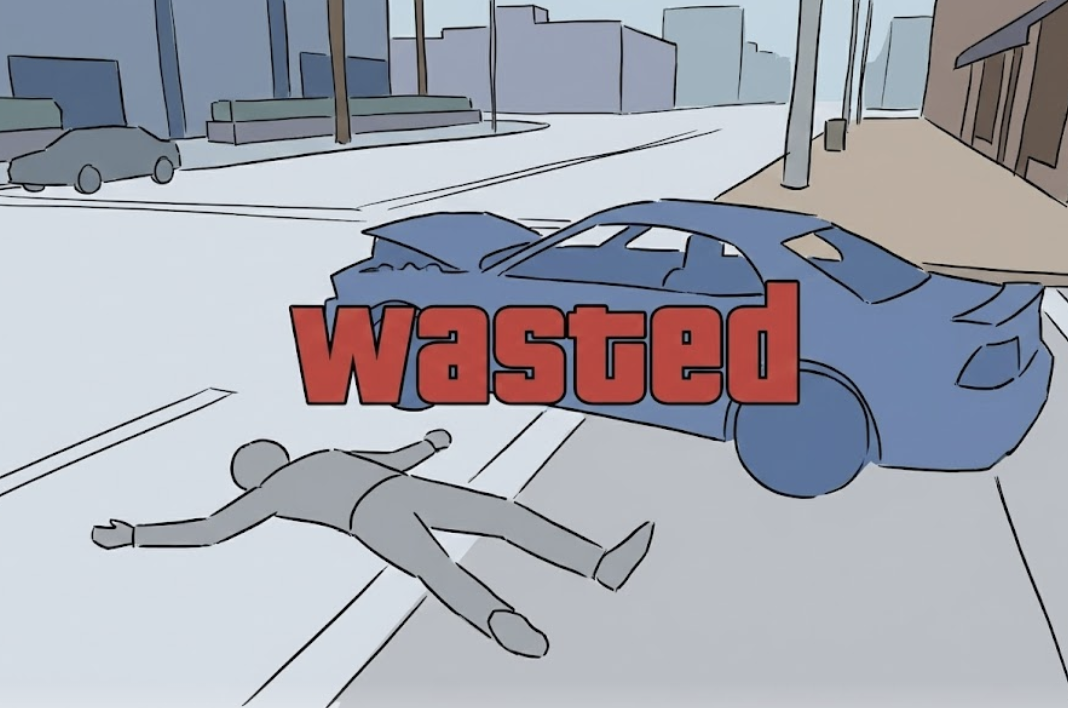}
\end{minipage}
\par\vspace{4pt}

\textbf{\textcolor{black}{Summary:}}
The game over sound effect in Grand Theft Auto. \\[3pt]
\textbf{\textcolor{black}{Usage:}}
Used when someone fails something. \\
\textbf{\textcolor{black}{Emotion:}}
neutral \\
\textbf{\textcolor{black}{Sensitivity:}}
none \\
\textbf{\textcolor{black}{Question:}}
\emph{Where will you likely be after you hear this sound?}
\begin{enumerate}[label=\textbf{\Alph*.}, nosep, leftmargin=*]
    \item Home.
    \item \textbf{Hospital.}
    \item Police office.
    \item Jail.
    \item Another world.
    \item Flight landing.
\end{enumerate}
\vspace{3pt}
\textbf{\textcolor{black}{Question Type:}} World Knowledge
\end{memeexample}
\end{minipage}
\hfill
\begin{minipage}[t]{0.48\textwidth}
\begin{memeexample}[colback=gray!2,colframe=light-red,
  title={\begin{CJK}{UTF8}{gbsn}杀死那个石家庄人 (卡祖笛版本)\end{CJK}}, left=4pt,right=4pt,top=4pt,bottom=4pt]
\small
\color{black!70}

\noindent
\begin{minipage}[t]{0.58\linewidth}
    \vspace{0pt}
    \textbf{\textcolor{black}{Category:}} music \\
    \textbf{\textcolor{black}{Year:}} 2010 \\
    \textbf{\textcolor{black}{Language:}} N/A \\
    \textbf{\textcolor{black}{Transcription:}} N/A
\end{minipage}%
\hfill
\begin{minipage}[t]{0.32\linewidth}
    \vspace{0pt}
    \raggedleft
    \includegraphics[width=\linewidth, keepaspectratio]{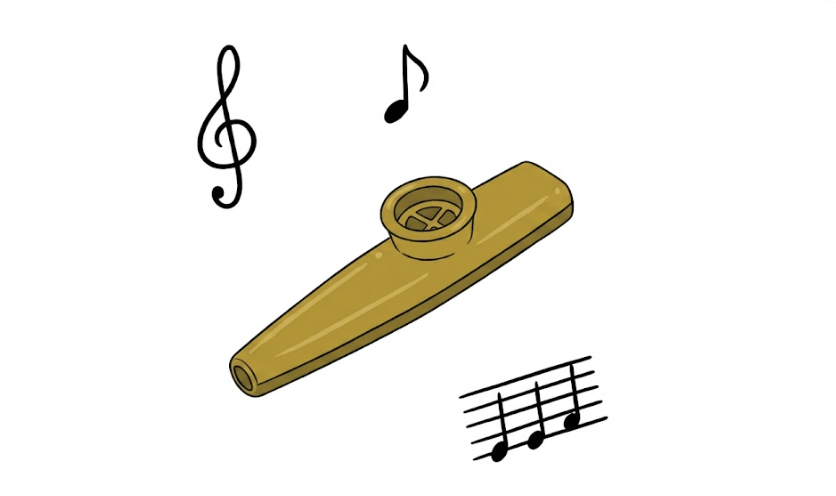}
\end{minipage}
\par\vspace{4pt}

\textbf{\textcolor{black}{Summary:}}
A short instrumental break mixing distorted guitar and kazoo, creating an ironic, mock-heroic feeling in the middle of a melancholic Chinese rock song. \\[3pt]
\textbf{\textcolor{black}{Usage:}}
Used to express absurd humor, irony, or mock-solemn moments in Chinese internet videos and memes. \\
\textbf{\textcolor{black}{Emotion:}}
sarcastic/ironic, nostalgic, sad \\
\textbf{\textcolor{black}{Sensitivity:}}
none \\
\textbf{\textcolor{black}{Question:}}
\emph{What type of character or figure would most likely appear when this distinctive instrumental interlude plays?}
\begin{enumerate}[label=\textbf{\Alph*.}, nosep, leftmargin=*]
    \item A cartoon superhero saving the world in slow motion with exaggerated seriousness.
    \item \textbf{A middle-aged man performing mundane tasks.}
    \item A young couple celebrating a wedding under bright fireworks.
    \item A musician playing a kazoo with a jazz song.
\end{enumerate}
\vspace{3pt}
\textbf{\textcolor{black}{Question Type:}} World Knowledge
\end{memeexample}
\end{minipage}

\noindent
\textbf{\textcolor{gemini-blue}{Gemini 3 Flash:}}
\underline{Jail}. \textcolor{red}{\xmark}\\

\noindent
\textbf{\textcolor{gemini-blue}{Gemini 3 Pro (low thinking):}}
\textcolor{gray}{\footnotesize
<think>**My Initial Analysis of the Sound**
Okay, so the user wants me to pinpoint the typical context of this sound effect and what it usually means. That's interesting. My first instinct is to break down the sound itself. It sounds like it kicks off with a really sharp, heavy impact – think slamming a heavy door or a metal gate slamming shut. That initial ``thud'' really sets the stage. Then, there's that metallic ringing or clanking that follows. It's almost like the sound of metal bars shifting or settling after the initial impact. It's a very distinctive sound, almost archetypal in my experience, and that's the key to identifying its typical use and cultural significance.
<$\backslash$think>
}
\underline{Another world}.\textcolor{red}{\xmark}\\

\noindent
\textbf{\textcolor{gemini-blue}{Gemini 3 Pro (high thinking):}}
\textcolor{gray}{\footnotesize
<think>**My Game Sense Kicked In**
Okay, first things first, I need to figure out where we're at based on this audio clip. *Thud, hum... wait a second...* That's it! That's the *Wasted* sound effect from GTA 5! That low, ominous drone is unmistakable. Instantly, the game mechanics flood my brain. When you get ``wasted'' in GTA 5, the screen goes grey, the big red letters pop up, and this glorious sound confirms your digital demise. So, where do you respawn? Not in some metaphorical ``other world,'' and definitely not at home unless you saved the game. Definitely not at the police office or jail; those are for getting ``busted,'' not ``wasted.'' Flight landing? Nope, no connection there. Ah! The hospital. That's it. Every time you die in GTA 5, they slap you with a medical bill and you wake up outside the nearest hospital.
<$\backslash$think>
}
\underline{Hospital}.\textcolor{green}{\cmark}\\

\noindent
\textbf{\textcolor{gemini-blue}{Gemini 3 Flash:}}\\
\underline{A middle-aged man performing mundane tasks.}\textcolor{green}{\cmark}\\

\noindent
\textbf{\textcolor{gemini-blue}{Gemini 3 Pro (low thinking):}}
\textcolor{gray}{\footnotesize
<think>**Analysis of a Comedic Audio Cue**
Alright, let's break this down. We've got a video with some music, and we need to figure out what kind of character or situation would *most likely* be associated with it. The key is the audio. It's a kazoo-like rendition of ``The Final Countdown'' – instantly recognizable, but intentionally goofy. My thought process immediately went to the contrast. ``The Final Countdown'' is an epic, dramatic song. But here, it's being played on a ridiculous instrument. The best fit is a scenario that mirrors this incongruity, this playful subversion of expectations. Option A, ``A cartoon superhero saving the world in slow motion with exaggerated seriousness,'' perfectly encapsulates this mock-epic vibe. It's that classic comedic juxtaposition: a serious situation rendered absurd by silly sound effects. We see this all the time in memes and parodies. This is where internet culture comes in. The sound itself is part of a *specific* meme format! Option B, a middle-aged man doing mundane tasks, is a common ironic choice for music but misses the specific energy of this song choice. It's a great example, but ``The Final Countdown'' has an inherent sense of building to something, so a middle-aged man washing dishes doesn't fit the 'countdown' or high-energy vibe. Option C is a straight-faced application of the audio, and there's nothing sincere about a kazoo playing at a wedding. Finally, Option D, while the video has a ``musician'' in it, the style is wrong (it's rock and not jazz). It's all about the tone! The goal is to maximize the humorous dissonance between the audio and the visual. If this *were* a question on a music or meme culture test, I'd say that there is only one correct answer - it has to be A. The ``cartoon superhero with exaggerated seriousness'' provides an immediate sense of the *mock-heroic* nature of the track. If the music *is* a key indicator, as is often the case with musical or meme questions, then A is the only choice that makes sense. Therefore, the answer must be A.
<$\backslash$think>
}
\underline{A cartoon superhero saving the world in slow} \underline{motion with exaggerated seriousness.}.\textcolor{red}{\xmark}\\

\noindent
\textbf{\textcolor{gemini-blue}{Gemini 3 Pro (high thinking):}}
\textcolor{gray}{\footnotesize
<think>**Alright, let's break this down.** The video's playing, and my mind's immediately going: ``That's *Shooting Stars* by Bag Raiders! But, wait... it's a kazoo?! And... strumming? Ok, let's analyze the question, the options, and the context here.'' My first thought? ``This is a meme reference, but which one?'' I rule out the basic \"falling through space\" visuals because the question's about the instrumental interlude. I quickly scan the options.  Superheroes don't quite fit the meme's usual silliness. Middle-aged man?  Nope, wrong vibe. Couple at a wedding? Too romantic. Then, it hits me: *Musician playing a kazoo*. Bingo! That's what the video is *literally* showing. Though there's a slight pause,  ``jazz song''? *Shooting Stars* isn't jazz, but maybe the question author is misinterpreting the solo nature of the kazoo as ``jazzy'' or just using that as a descriptor.I check myself, re-listening, and making sure the song is actually *Shooting Stars* by comparing the melody and the notes. It is!  Double-checking for other meme possibilities, but it's that kazoo cover!  The question's phrasing, ``What type of character or figure would most likely appear...'' is a bit ambiguous. However, I’m confident that it points to a literal description, and the best literal match is definitely D, even if the ``jazz'' bit is slightly off. The other options are distractors that do not relate to the video content.  My logic? The video *shows* a musician playing a kazoo, even if the genre isn't quite right.  That kazoo is the key here.  Ultimately, this boils down to which option best describes the video content. And that's D, no question.
<$\backslash$think>
}
\underline{A musician playing a kazoo with a jazz song.}\textcolor{red}{\xmark}\\

%\subsection{Human Results Analysis}
%\label{app:human_results}

\newpage

\section{More AVMeme Samples}
\label{app:examples}
\begin{multicols}{2}

% father

\begin{memeexample}[colback=gray!2,colframe=light-purple,
  title={I am your father.}, left=4pt,right=4pt,top=4pt,bottom=4pt]
% 1. Set font to small
\small
% 2. Set general text color to dark gray (70% black)
\color{black!70}

% --- Start of image insertion block ---
\noindent
\begin{minipage}[t]{0.5\linewidth}
    \vspace{0pt}
    \textbf{\textcolor{black}{Category:}} noisy speech \\
    \textbf{\textcolor{black}{Year:}} 1980 \\
    \textbf{\textcolor{black}{Language:}} English \\
\end{minipage}%
\hfill
\begin{minipage}[t]{0.35\linewidth}
    \vspace{0pt}
    \raggedleft
    % Placeholder image as requested
    \includegraphics[width=\linewidth, keepaspectratio]{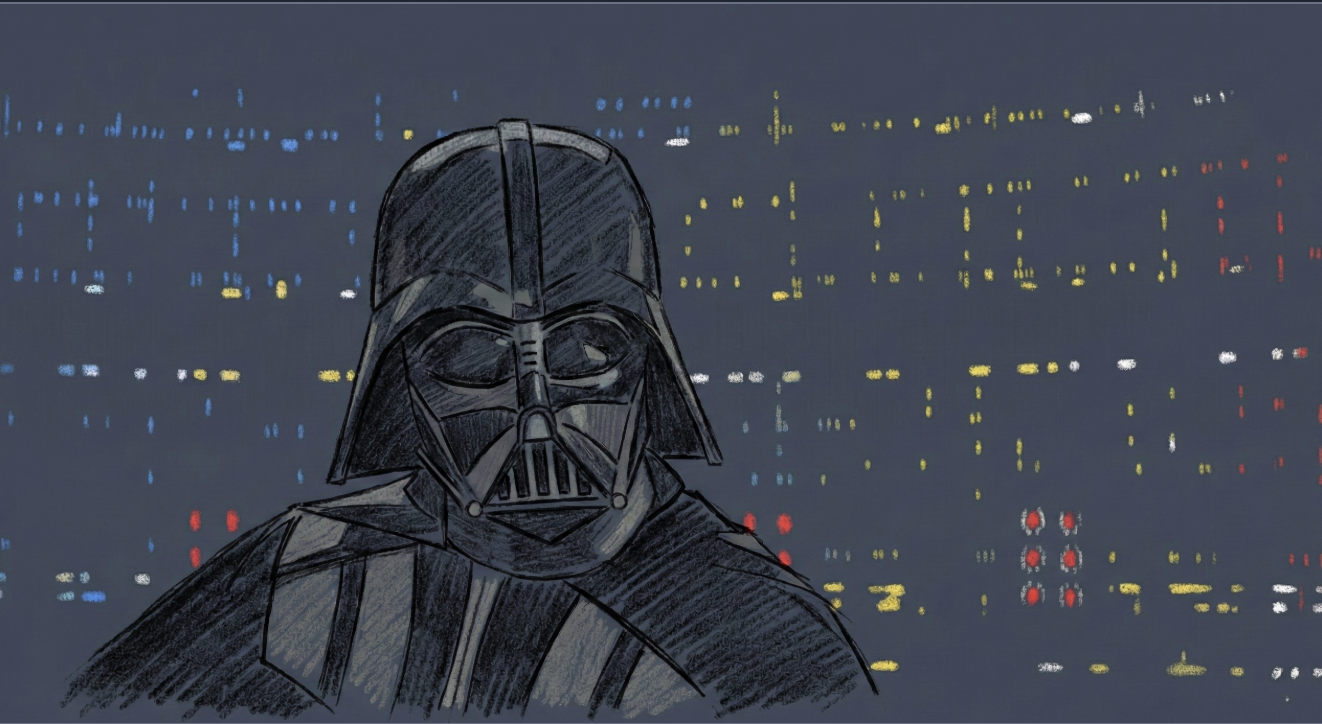}
\end{minipage}
\par\vspace{4pt}
% --- End of image insertion block ---
\begin{CJK}{UTF8}{min}
\textbf{\textcolor{black}{Transcription:}} Luke: He told me you killed him. Vader: No. I am your father. Luke: No. No. It's not true.
\end{CJK}\\
\textbf{\textcolor{black}{Summary:}} A man reveals to a younger man that he is his father.
 \\[3pt]
\textbf{\textcolor{black}{Usage:}} Used as a parody of the famous reveal scene, often remixed as “No, I am your X.”
 \\
\textbf{\textcolor{black}{Emotion:}} sad, surprised/shocked
 \\
\textbf{\textcolor{black}{Sensitivity:}} violence/gore
 \\
\textbf{\textcolor{black}{Question:}}
\emph{What will the man most likely do after he knows the truth?}
\begin{enumerate}[label=\textbf{\Alph*.}, nosep, leftmargin=*]
    \item Fight.
    \item \textbf{Escape.}
    \item Surrender.
    \item Stay put.
\end{enumerate}
\textbf{\textcolor{black}{Question Type:}} Contextual Inference
\end{memeexample}

% win

\begin{memeexample}[colback=gray!2,colframe=light-blue,
  title={Windows Error}, left=4pt,right=4pt,top=4pt,bottom=4pt]
% 1. Set font to small
\small
% 2. Set general text color to dark gray (70% black)
\color{black!70}

% --- Start of image insertion block ---
\noindent
\begin{minipage}[t]{0.5\linewidth}
    \vspace{0pt}
    \textbf{\textcolor{black}{Category:}} sound effect \\
    \textbf{\textcolor{black}{Year:}} 2001 \\
    \textbf{\textcolor{black}{Language:}} N/A \\
\end{minipage}%
\hfill
\begin{minipage}[t]{0.35\linewidth}
    \vspace{0pt}
    \raggedleft
    % Placeholder image as requested
    \includegraphics[width=\linewidth, keepaspectratio]{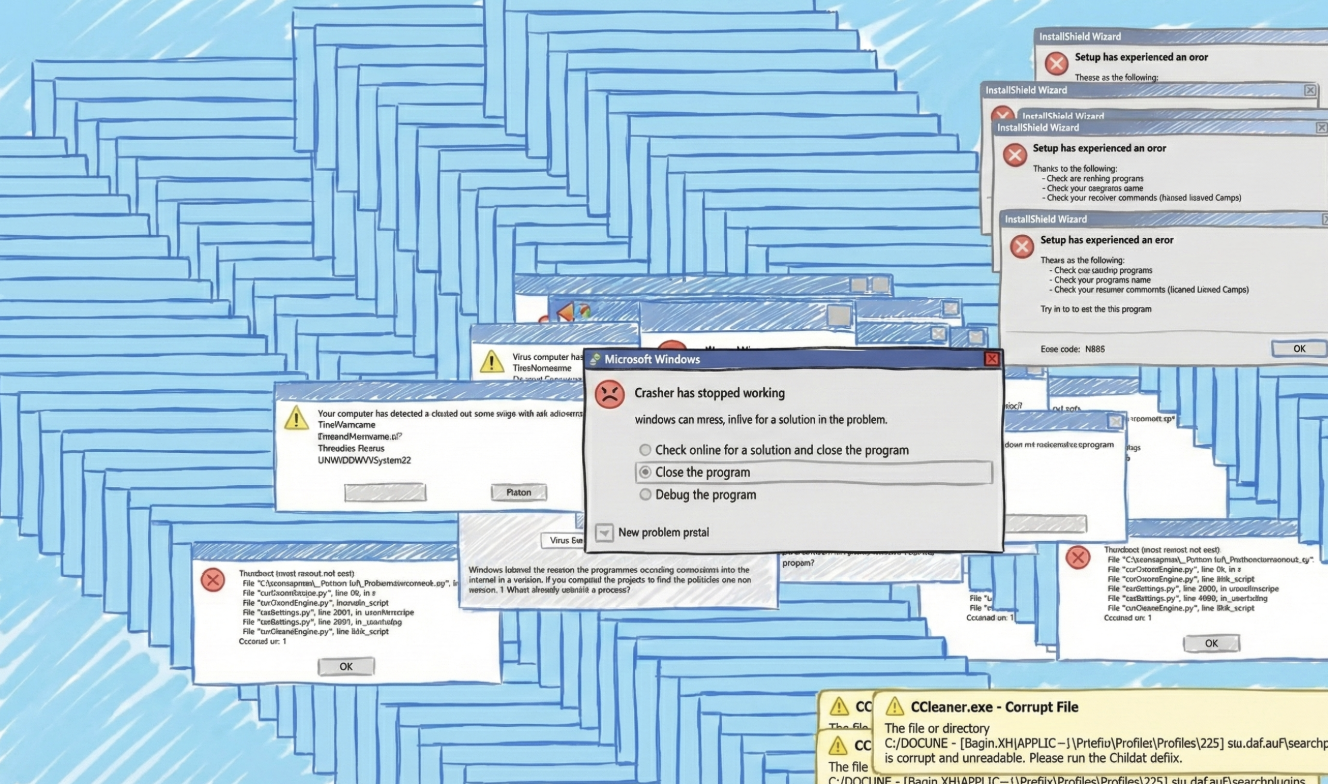}
\end{minipage}
% --- End of image insertion block ---
\begin{CJK}{UTF8}{min}
\textbf{\textcolor{black}{Transcription:}} N/A
\end{CJK}\\
\textbf{\textcolor{black}{Summary:}} Infinite pop-up errors in Windows.
 \\[3pt]
\textbf{\textcolor{black}{Usage:}} The sound effect of the error is widely recognized and used for comic “fail” beats
 \\
\textbf{\textcolor{black}{Emotion:}} angry/annoyed
 \\
\textbf{\textcolor{black}{Sensitivity:}} none
 \\
\textbf{\textcolor{black}{Question:}}
\emph{Which operating system makes this sound?}
\begin{enumerate}[label=\textbf{\Alph*.}, nosep, leftmargin=*]
    \item \textbf{Windows}
    \item macOS
    \item Linux
    \item Android
    \item iOS
\end{enumerate}
\textbf{\textcolor{black}{Question Type:}} World Knowledge \\
\textbf{\textcolor{black}{visual\_cheat: }\textcolor{red}{True}}
\end{memeexample}

\begin{memeexample}[colback=gray!2,colframe=light-pink,
  title={Liszt's Hungarian Rhapsody No.2}, left=4pt,right=4pt,top=4pt,bottom=4pt]
% 1. Set font to small
\small
% 2. Set general text color to dark gray (70% black)
\color{black!70}

% --- Start of image insertion block ---
\noindent
\begin{minipage}[t]{0.5\linewidth}
    \vspace{0pt}
    \textbf{\textcolor{black}{Category:}} music \\
    \textbf{\textcolor{black}{Year:}} 1851 \\
    \textbf{\textcolor{black}{Language:}} N/A \\
    \textbf{\textcolor{black}{Transcription:}} N/A
\end{minipage}%
\hfill
\begin{minipage}[t]{0.35\linewidth}
    \vspace{0pt}
    \raggedleft
    % Placeholder image as requested
    \includegraphics[width=\linewidth, keepaspectratio]{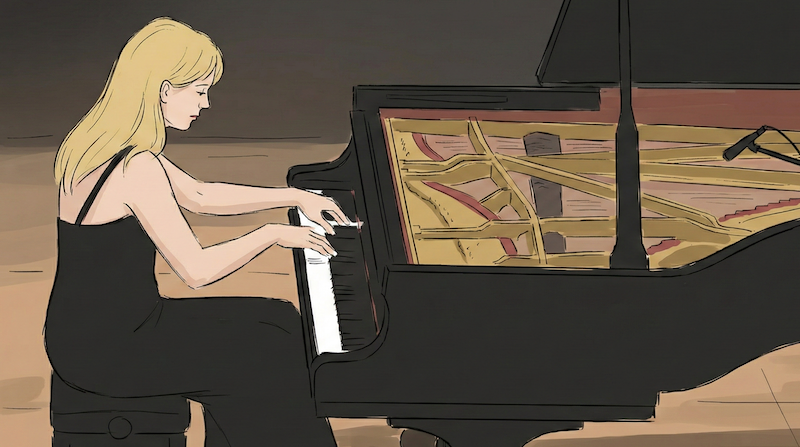}
\end{minipage}
\par\vspace{4pt}
% --- End of image insertion block ---

\textbf{\textcolor{black}{Summary:}}
A fast playful dance-like solo piano music. \\[3pt]
\textbf{\textcolor{black}{Usage:}}
Used as the background music often in cartoons for chase and slapstick scenes. \\
\textbf{\textcolor{black}{Emotion:}}
happy \\
\textbf{\textcolor{black}{Sensitivity:}}
none \\
\textbf{\textcolor{black}{Question:}}
\emph{On which of the following occasions will you most likely hear this music?}
\begin{enumerate}[label=\textbf{\Alph*.}, nosep, leftmargin=*]
    \item TV ads.
    \item \textbf{Cartoon.}
    \item Daily news.
    \item Talk show.
\end{enumerate}
\textbf{\textcolor{black}{Question Type:}} Usage \& Application
\end{memeexample}

\begin{memeexample}[colback=gray!2,colframe=light-green,
  title={Start digging in your butt twin}, left=4pt,right=4pt,top=4pt,bottom=4pt]
% 1. Set font to small
\small
% 2. Set general text color to dark gray (70% black)
\color{black!70}

% --- Start of image insertion block ---
\noindent
\begin{minipage}[t]{0.5\linewidth}
    \vspace{0pt}
    \textbf{\textcolor{black}{Category:}} clean speech \\
    \textbf{\textcolor{black}{Year:}} 2025 \\
    \textbf{\textcolor{black}{Language:}} en \\
    \textbf{\textcolor{black}{Transcription:}} A: And I want you to make me a Krabby Patty. B: Start diggin' in your butt twin. A: What? Wha--
\end{minipage}%
\hfill
\begin{minipage}[t]{0.35\linewidth}
    \vspace{0pt}
    \raggedleft
    % Placeholder image as requested
    \includegraphics[width=\linewidth, keepaspectratio]{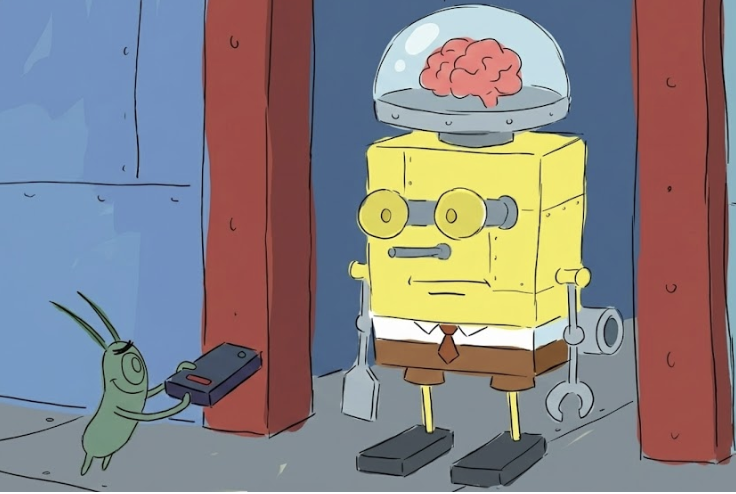}
\end{minipage}
% --- End of image insertion block ---

\textbf{\textcolor{black}{Summary:}}
A crude absurdist catchphrase from TikTok, telling someone to “start digging in your butt, twin.” Often used with SpongeBob edits or as nonsense humor. \\[3pt]
\textbf{\textcolor{black}{Usage:}}
Used in TikTok edits or friend chats as absurd humor. Sometimes implies “don't be uptight” or “stop talking nonsense,” but often it's simply meaningless “brain-rot” humor that exaggerates absurdity. \\
\textbf{\textcolor{black}{Emotion:}}
sarcastic/ironic, cringe/embarrassed\\
\textbf{\textcolor{black}{Sensitivity:}}
sex \\
\textbf{\textcolor{black}{Question:}}
\emph{What is the most likely purpose of this clip?}
\begin{enumerate}[label=\textbf{\Alph*.}, nosep, leftmargin=*]
    \item Seeking clarity.
    \item \textbf{Nudge playfully.}
    \item Warn seriously.
    \item Request support.
\end{enumerate}
\textbf{\textcolor{black}{Question Type:}} Usage \& Application
\end{memeexample}

% Miku

\begin{memeexample}[colback=gray!2,colframe=light-cyan,
  title={Miku Hatsune's Ievan Polkka}, left=4pt,right=4pt,top=4pt,bottom=4pt]
% 1. Set font to small
\small
% 2. Set general text color to dark gray (70% black)
\color{black!70}

% --- Start of image insertion block ---
\noindent
\begin{minipage}[t]{0.5\linewidth}
    \vspace{0pt}
    \textbf{\textcolor{black}{Category:}} song \\
    \textbf{\textcolor{black}{Year:}} 2006 \\
    \textbf{\textcolor{black}{Language:}} Finnish \\
\end{minipage}%
\hfill
\begin{minipage}[t]{0.35\linewidth}
    \vspace{0pt}
    \raggedleft
    % Placeholder image as requested
    \includegraphics[width=\linewidth, keepaspectratio]{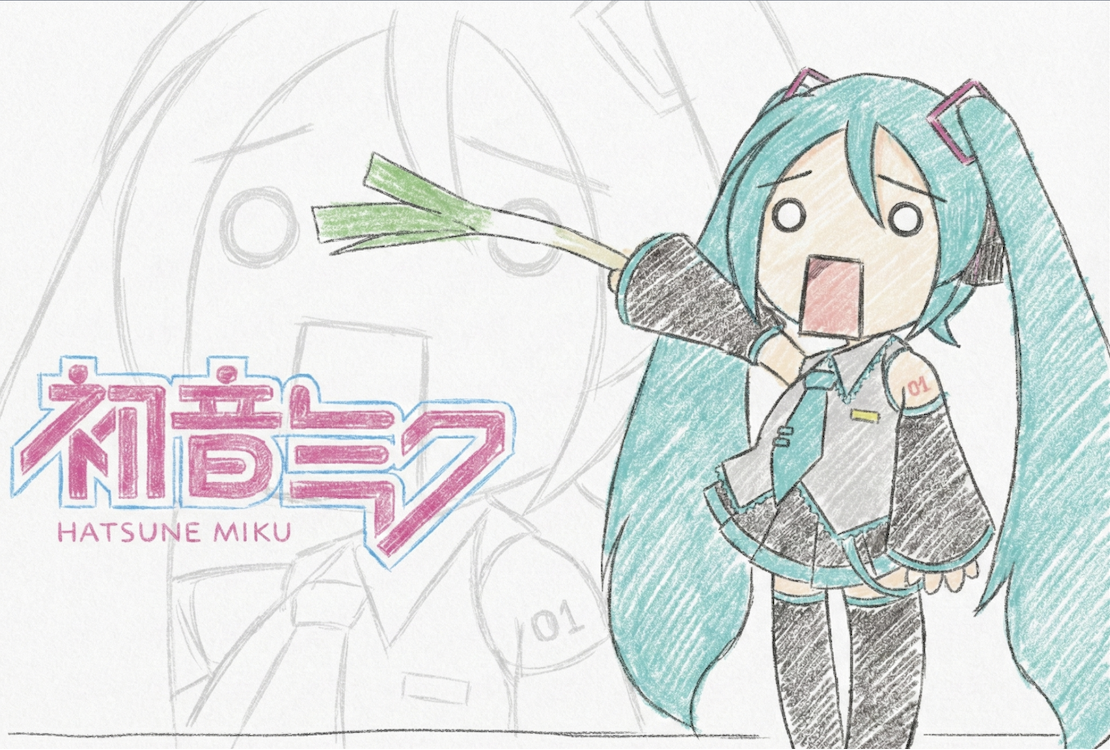}
\end{minipage}
\par\vspace{4pt}
% --- End of image insertion block ---
\begin{CJK}{UTF8}{min}
\textbf{\textcolor{black}{Transcription:}} Ratsatsaa ja ripidabi dilla, Beritstan dillan dellan doo, A baribbattaa baribbariiba, Ribiribi distan dellan doo, Ja barillas dillan deia dooa, Daba daba daba daba daba duvja vuu.
\end{CJK}\\
\textbf{\textcolor{black}{Summary:}} The famous Leekspin originated from the Finnish song “Ievan Polkka,” sung by the Finnish quartet Loituma. It first went viral after being paired with a looping animation of Orihime Inoue twirling a leek from the anime Bleach. Later, it became even more popular thanks to the Hatsune Miku version, along with the creation of the “Loituma Girl” (Negi Miku) character based on her.
 \\[3pt]
\textbf{\textcolor{black}{Usage:}} Used for cover performances, dance routines, fan remixes (such as other anime characters spinning leeks), and as background music in cheerful or speed up videos.
 \\
\textbf{\textcolor{black}{Emotion:}} happy
 \\
\textbf{\textcolor{black}{Sensitivity:}} none
 \\
\textbf{\textcolor{black}{Question:}}
\emph{Who provided the voice for the first ACG character associated with this song?}
\begin{enumerate}[label=\textbf{\Alph*.}, nosep, leftmargin=*]
    \item Saki Fujita.
    \item \textbf{Yuki Matsuoka.}
    \item Rina Satō.
    \item Fumiko Orikasa.
\end{enumerate}
\vspace{3pt}
\textbf{\textcolor{black}{Question Type:}} World Knowledge
\end{memeexample}

\begin{memeexample}[colback=gray!2,colframe=light-orange,
  title={I don't like sand}, left=4pt,right=4pt,top=4pt,bottom=4pt]
% 1. Set font to small
\small
% 2. Set general text color to dark gray (70% black)
\color{black!70}
% --- Start of image insertion block ---
\noindent % Ensures the block starts at the left margin
% First minipage for the text fields (takes up ~68% of width)
\begin{minipage}[t]{0.6\linewidth}
    \vspace{0pt} % This invisible anchor helps with top-alignment
    \textbf{\textcolor{black}{Category:}} noisy speech \\
    \textbf{\textcolor{black}{Year:}} 2002 \\
    \textbf{\textcolor{black}{Language:}} English \\
    \textbf{\textcolor{black}{Transcription:}} I don't like sand. 
\end{minipage}%
\hfill % Fills the space between, pushing the image right
% Second minipage for the image (takes up ~30% of width)
\begin{minipage}[t]{0.3\linewidth}
    \vspace{0pt} % Anchor for top-alignment
    \raggedleft % Aligns the image to the right inside its mini-box
    % Insert your image here. Adjust 'width' to control the "stamp" size.
    % Replace 'image_0.png' with your actual image filename.
    \includegraphics[width=\linewidth, keepaspectratio]{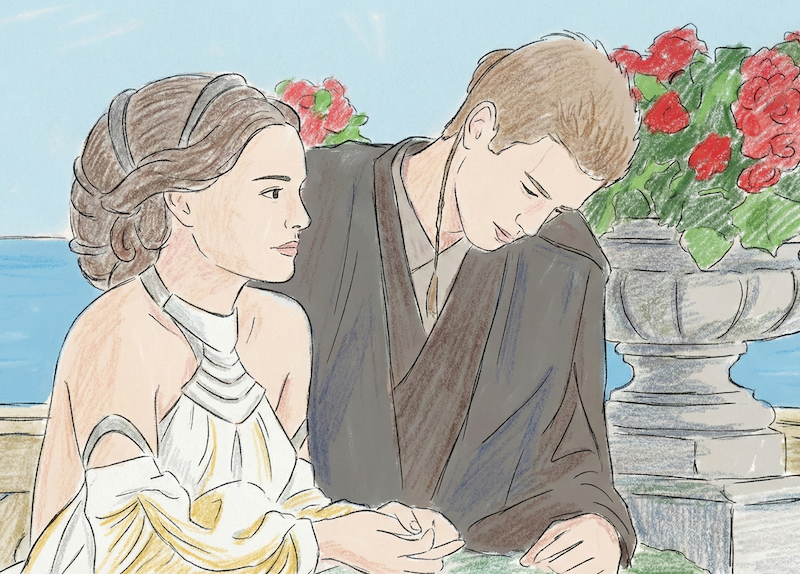}
\end{minipage}
\textbf{\textcolor{black}{Summary:}}
Anakin Skywalker, in an awkward attempt at flirting, complains about sand while comparing it to the person he is with. \\[3pt]
\textbf{\textcolor{black}{Usage:}}
Used to mock awkward dialogue, complain about minor inconveniences in a melodramatic way, or as a general reference to the Star Wars prequel trilogy. \\
\textbf{\textcolor{black}{Emotion:}}
cringe/embarrassed \\
\textbf{\textcolor{black}{Sensitivity:}}
none \\
\textbf{\textcolor{black}{Question:}}
\emph{What is the speaker’s primary motivation for making this comment about sand?}
\begin{enumerate}[label=\textbf{\Alph*.}, nosep, leftmargin=*]
    \item To provide a detailed geological analysis.
    \item To share a fond childhood memory.
    \item \textbf{As an awkward attempt to compliment his companion.}
    \item To change the subject to his favorite topic, desert planets.
\end{enumerate}
\textbf{\textcolor{black}{Question Type:}} Contextual Inference
\end{memeexample}

% Chicken

\begin{memeexample}[colback=gray!2,colframe=light-green,
  title={Look at all those chickens}, left=4pt,right=4pt,top=4pt,bottom=4pt]
% 1. Set font to small
\small
% 2. Set general text color to dark gray (70% black)
\color{black!70}

% --- Start of image insertion block ---
\noindent
\begin{minipage}[t]{0.5\linewidth}
    \vspace{0pt}
    \textbf{\textcolor{black}{Category:}} clean speech \\
    \textbf{\textcolor{black}{Year:}} 2013 \\
    \textbf{\textcolor{black}{Language:}} English \\
\end{minipage}%
\hfill
\begin{minipage}[t]{0.35\linewidth}
    \vspace{0pt}
    \raggedleft
    % Placeholder image as requested
    \includegraphics[width=\linewidth, keepaspectratio]{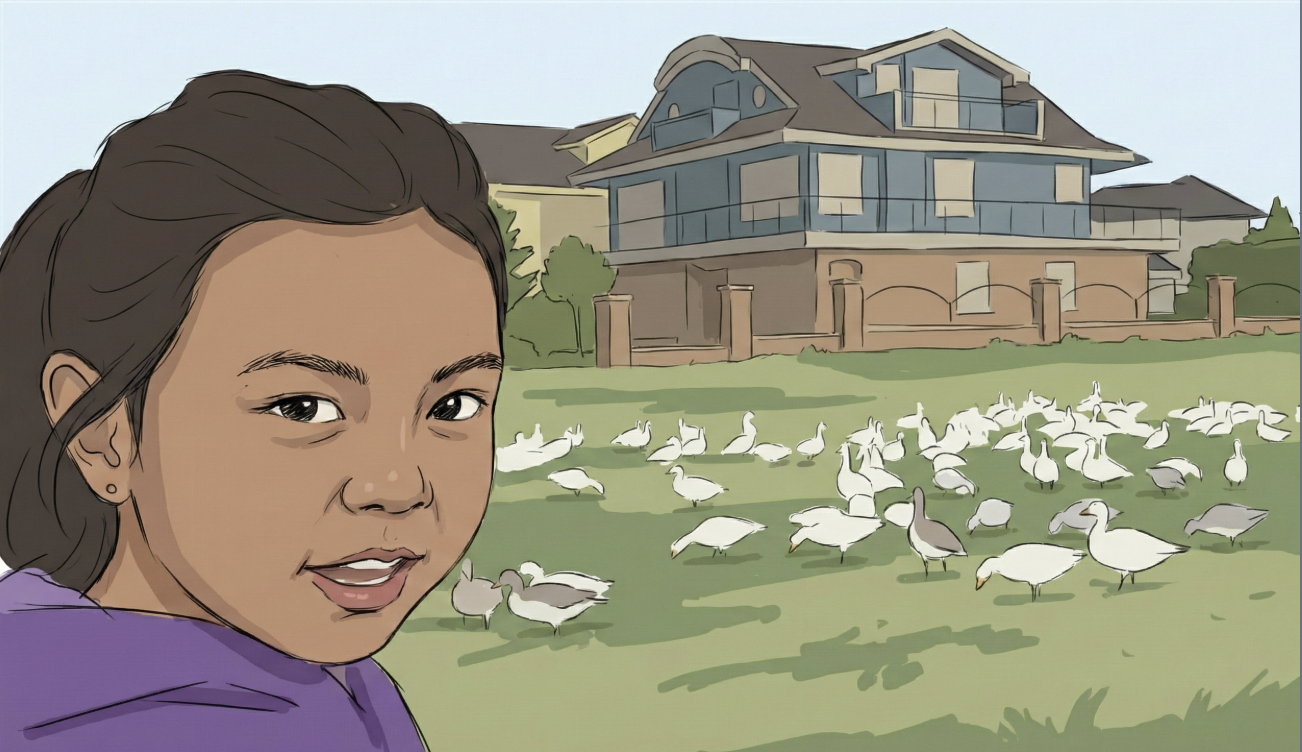}
\end{minipage}
\par\vspace{4pt}
% --- End of image insertion block ---
\begin{CJK}{UTF8}{min}
\textbf{\textcolor{black}{Transcription:}} Look at all those chickens
\end{CJK}\\
\textbf{\textcolor{black}{Summary:}} A young girl, awestruck, exclaims ``Look at all those chickens!`` while looking at a flock of what are actually geese or ducks.
 \\[3pt]
\textbf{\textcolor{black}{Usage:}} Used to express excitement or amazement at a large number of things, often while misidentifying them.
 \\
\textbf{\textcolor{black}{Emotion:}} happy, surprised/shocked
 \\
\textbf{\textcolor{black}{Sensitivity:}} none
 \\
\textbf{\textcolor{black}{Question:}}
\emph{What is the most likely reason for the person's exclamation?}
\begin{enumerate}[label=\textbf{\Alph*.}, nosep, leftmargin=*]
    \item She is scared of the animals.
    \item \textbf{She is excited and misidentifying the animals.}
    \item She is angry at the animals for making a mess.
    \item She is trying to sell the animals.
\end{enumerate}
\textbf{\textcolor{black}{Question Type:}} Contextual Inference
\end{memeexample}

% Duang

\begin{memeexample}[colback=gray!2,colframe=light-red,
  title={\begin{CJK}{UTF8}{gbsn}霸王洗发水\end{CJK}}, left=4pt,right=4pt,top=4pt,bottom=4pt]
% 1. Set font to small
\small
% 2. Set general text color to dark gray (70% black)
\color{black!70}

% --- Start of image insertion block ---
\noindent
\begin{minipage}[t]{0.5\linewidth}
    \vspace{0pt}
    \textbf{\textcolor{black}{Category:}} noisy speech \\
    \textbf{\textcolor{black}{Year:}} 2004 \\
    \textbf{\textcolor{black}{Language:}} Chinese \\
\end{minipage}%
\hfill
\begin{minipage}[t]{0.35\linewidth}
    \vspace{0pt}
    \raggedleft
    % Placeholder image as requested
    \includegraphics[width=\linewidth, keepaspectratio]{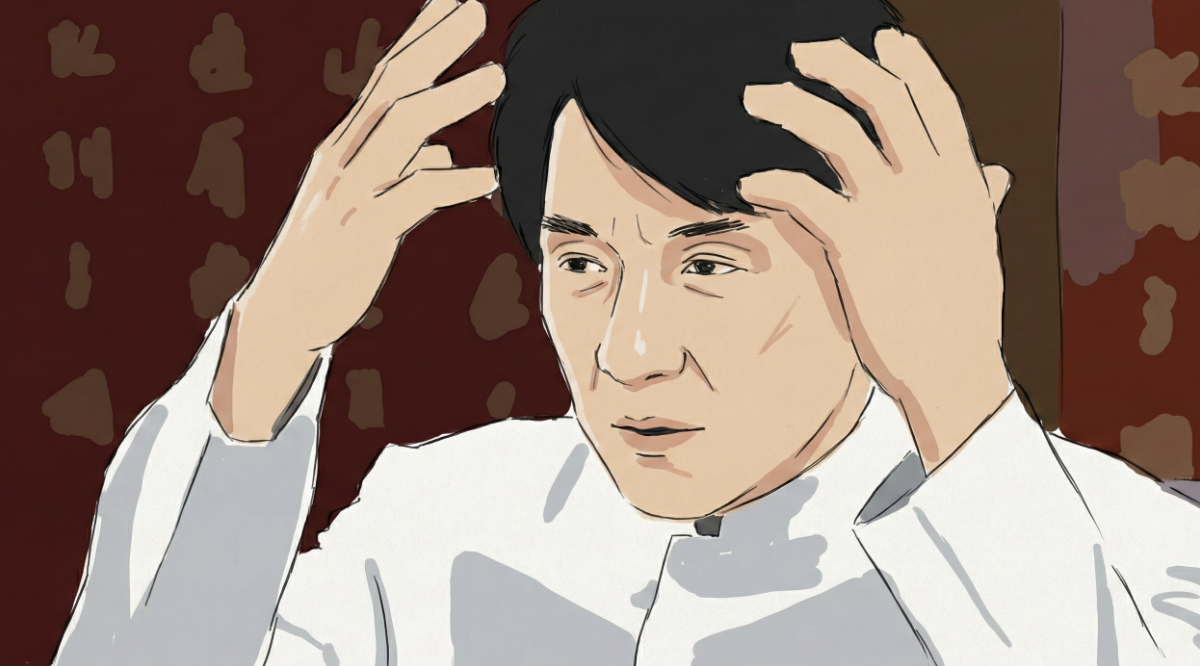}
\end{minipage}
\par\vspace{4pt}
% --- End of image insertion block ---
\begin{CJK}{UTF8}{gbsn}
\textbf{\textcolor{black}{Transcription:}} 当我第一次知道要拍那个洗发水广告的时候,其实我是拒绝的。因为我觉得,你不能叫我拍我就马上去拍,第一我要试一下,不然我又不想说,一个广告拍完后加上好多特技,那个头发,Duang,很黑,很亮,很柔,结果观众出来一定会骂我,根本没有这种头发,证明上面这个是假的。我说先给我试一下。后来我经过也知道它们是中药的,而且没有那种化学成分的。
\end{CJK}\\
\textbf{\textcolor{black}{Summary:}} An actor says he first refused to shoot the shampoo ad with the worry that special effects would make the hair look unreal, and only agreed after testing it.
 \\[3pt]
\textbf{\textcolor{black}{Usage:}} “Duang” is often used for a flashy, sudden, over-the-top sound or visual effect.
 \\
\textbf{\textcolor{black}{Emotion:}} neutral
 \\
\textbf{\textcolor{black}{Sensitivity:}} none
 \\
\textbf{\textcolor{black}{Question:}}
\emph{Which one of the following is the least suitable to be described by the onomatopoeic word in this clip?}
\begin{enumerate}[label=\textbf{\Alph*.}, nosep, leftmargin=*]
    \item Luxury race car.
    \item Silky-shiny hair.
    \item \textbf{Dramatic sea view.}
    \item New game with good visuals.
\end{enumerate}
\vspace{3pt}
\textbf{\textcolor{black}{Question Type:}} Usage \& Application
\end{memeexample}

\begin{memeexample}[colback=gray!2,colframe=light-blue,
  title={\begin{CJK}{UTF8}{gbsn}超威蓝猫\end{CJK}}, left=4pt,right=4pt,top=4pt,bottom=4pt]

% 1. Set font to small
\small
% 2. Set general text color to dark gray (70% black)
\color{black!70}

% --- Start of image insertion block ---
\noindent
\begin{minipage}[t]{0.5\linewidth}
    \vspace{0pt}
    \textbf{\textcolor{black}{Category:}} clean speech \\
    \textbf{\textcolor{black}{Year:}} 2009 \\
    \textbf{\textcolor{black}{Language:}} Chinese \\
\end{minipage}%
\hfill
\begin{minipage}[t]{0.35\linewidth}
    \vspace{0pt}
    \raggedleft
    % Placeholder image
    \includegraphics[width=\linewidth, keepaspectratio]{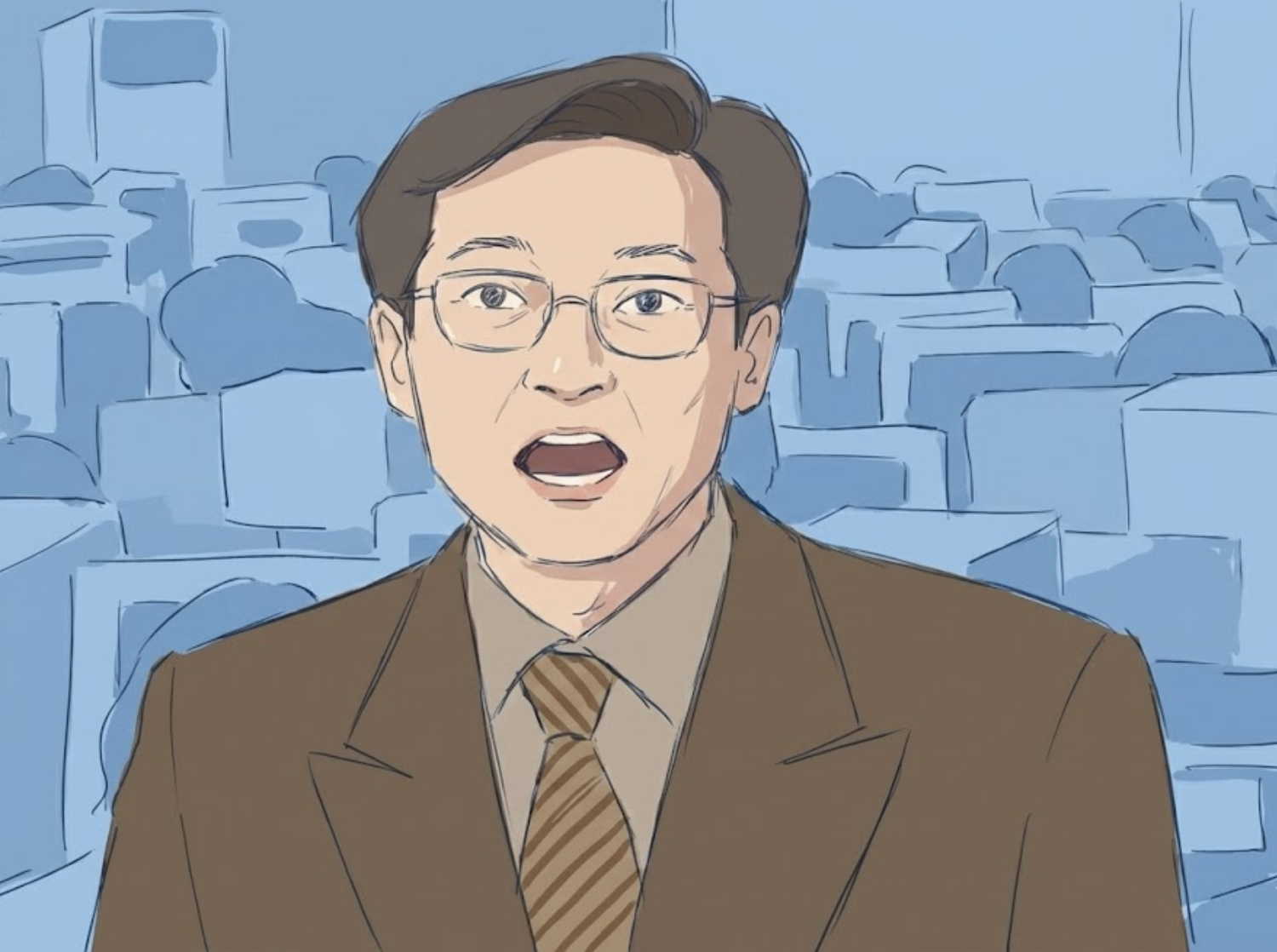}
\end{minipage}
\par\vspace{4pt}
% --- End of image insertion block ---

\begin{CJK}{UTF8}{gbsn}
\textbf{\textcolor{black}{Transcription:}}
我说小朋友们谁愿意跟我蓝猫学一句配音呐? 只听见下面有一个：我来！ 这时候我一看！ 小朋友们，人群之中钻出来一个光头！ 哇！他全身穿的都是我们蓝猫品牌的衣服，特别显眼。 我一把就把他抱了起来。
\end{CJK}\\

\textbf{\textcolor{black}{Summary:}}
The voice actor calls for a voice-dubbing volunteer, and a bald kid in full Blue Cat clothes eagerly shouts out.
\\[3pt]

\textbf{\textcolor{black}{Usage:}}
Used as a prosody when someone with a bald head suddenly appears.
\\

\textbf{\textcolor{black}{Emotion:}}
happy, surprised/shocked
\\

\textbf{\textcolor{black}{Sensitivity:}}
none
\\

\textbf{\textcolor{black}{Question:}}
\emph{What is the funniest thing here?}

\begin{enumerate}[label=\textbf{\Alph*.}, nosep, leftmargin=*]
    \item Cartoon suit.
    \item Shouting out.
    \item The voice acting.
    \item \textbf{Bare hair.}
\end{enumerate}

\vspace{3pt}
\textbf{\textcolor{black}{Question Type:}} Humor \& Popularity

\end{memeexample}

\begin{memeexample}[colback=gray!2,colframe=light-gray,
  title={\begin{CJK}{UTF8}{min}僕は新世界の神となる\end{CJK}}, left=4pt,right=4pt,top=4pt,bottom=4pt]
% 1. Set font to small
\small
% 2. Set general text color to dark gray (70% black)
\color{black!70}

% --- Start of image insertion block ---
\noindent
\begin{minipage}[t]{0.5\linewidth}
    \vspace{0pt}
    \textbf{\textcolor{black}{Category:}} clean speech \\
    \textbf{\textcolor{black}{Year:}} 2003 \\
    \textbf{\textcolor{black}{Language:}} Japanese \\
\end{minipage}%
\hfill
\begin{minipage}[t]{0.35\linewidth}
    \vspace{0pt}
    \raggedleft
    % Placeholder image as requested
    \includegraphics[width=\linewidth, keepaspectratio]{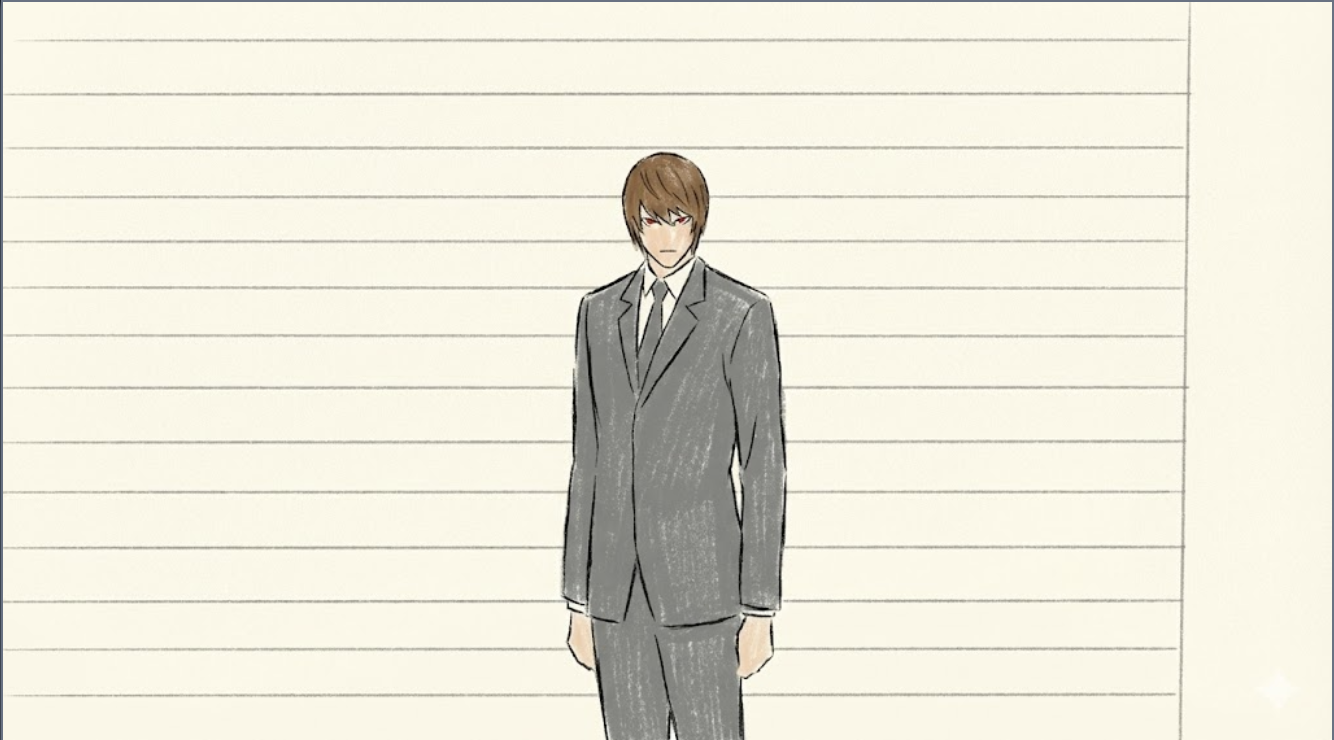}
\end{minipage}
\par\vspace{4pt}
% --- End of image insertion block ---
\begin{CJK}{UTF8}{min}
\textbf{\textcolor{black}{Transcription:}} いいか、僕はキラ。\\そして、新世界の神！
\end{CJK}\\
\textbf{\textcolor{black}{Summary:}}
A man claims he is God. \\[3pt]
\textbf{\textcolor{black}{Usage:}}
Used as a way of parodying Light's god complex by dropping his serious line into completely unserious, usually funny situations. \\
\textbf{\textcolor{black}{Emotion:}}
nostalgic, sarcastic/ironic, cringe/embarrassed \\
\textbf{\textcolor{black}{Sensitivity:}}
none \\
\textbf{\textcolor{black}{Question:}}
\emph{Why is he claiming that he is a god?}
\begin{enumerate}[label=\textbf{\Alph*.}, nosep, leftmargin=*]
    \item Because he is a god.
    \item \textbf{Because he has supernatural power.}
    \item Because he is mentally ill and delusional.
\end{enumerate}
\textbf{\textcolor{black}{Question Type:}} Contextual Inference
\end{memeexample}

% Ryutaro Nonomura

\begin{memeexample}[colback=gray!2,colframe=light-orange,
  title={\begin{CJK}{UTF8}{min}野々村議員\end{CJK}}, left=4pt,right=4pt,top=4pt,bottom=4pt]
% 1. Set font to small
\small
% 2. Set general text color to dark gray (70% black)
\color{black!70}

% --- Start of image insertion block ---
\noindent
\begin{minipage}[t]{0.5\linewidth}
    \vspace{0pt}
    \textbf{\textcolor{black}{Category:}} noisy speech \\
    \textbf{\textcolor{black}{Year:}} 2014 \\
    \textbf{\textcolor{black}{Language:}} Japanese \\
\end{minipage}%
\hfill
\begin{minipage}[t]{0.35\linewidth}
    \vspace{0pt}
    \raggedleft
    % Placeholder image as requested
    \includegraphics[width=\linewidth, keepaspectratio]{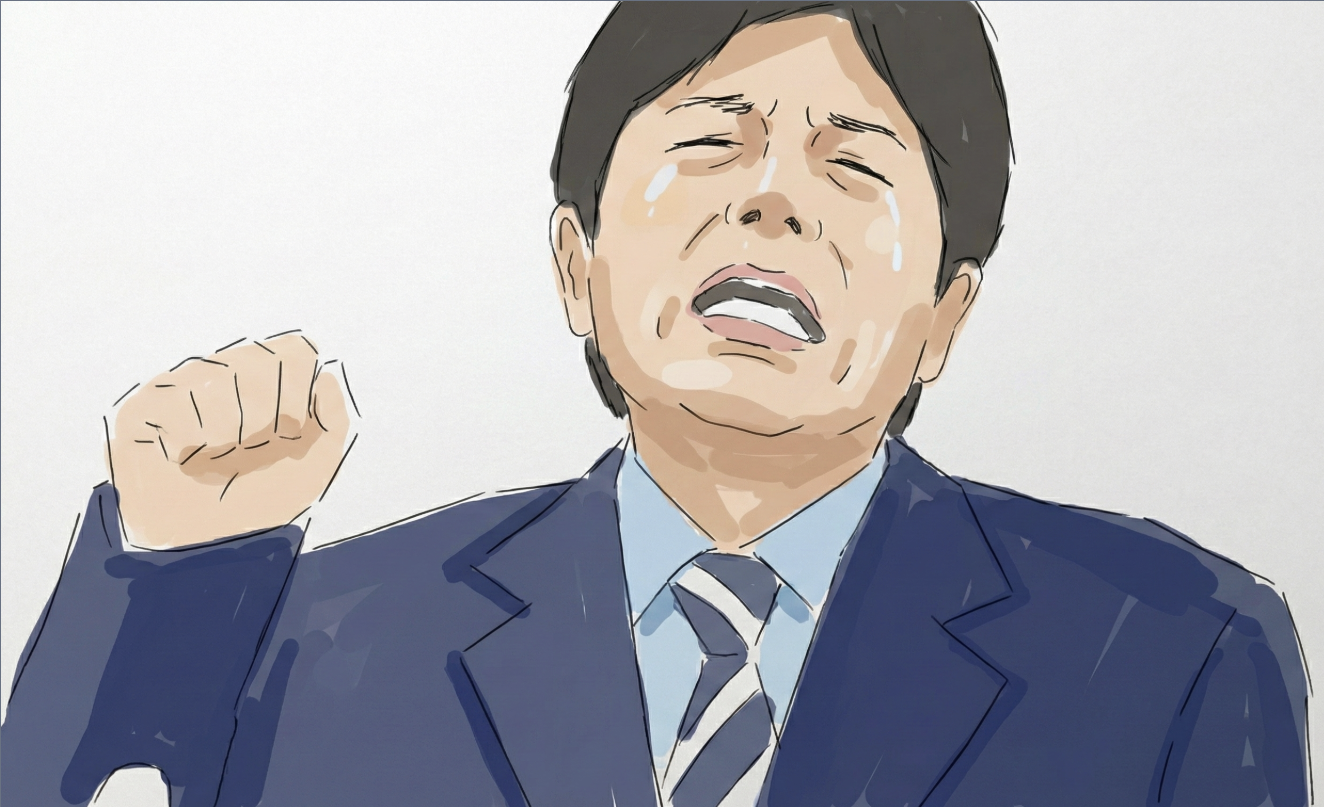}
\end{minipage}
\par\vspace{4pt}
% --- End of image insertion block ---
\begin{CJK}{UTF8}{min}
\textbf{\textcolor{black}{Transcription:}} この世の中を! ウグッブーン!! ゴノ ゴノ世の ブッヒィフエエエーーーンン!! ヒィェーーッフウンン!! ウゥ……ウゥ…… ア゛ーーーーー ア゛ッア゛ーー!!! ゴノぉよぉっきゃガッハッハアン!! ア゛ーー 世の中を! うぅ変エダイ! その一心でえぇ!! ィヒー フーッ ハゥ 一生懸命訴えて 西宮\$¥\#\%\& 縁もゆかりもない 西宮ッヘエ市民の 皆さまに選出されて! やっと! 議員に! なったんですうぅー!!!
\end{CJK}\\
\textbf{\textcolor{black}{Summary:}} Mr.Nonomura is holding a press conference in which he is expressing his apology for misusing publidc funds.
 \\[3pt]
\textbf{\textcolor{black}{Usage:}} Used as a reaction image to express fake or exaggerated sadness
 \\
\textbf{\textcolor{black}{Emotion:}} cringe/embarrassed
 \\
\textbf{\textcolor{black}{Sensitivity:}} none
 \\
\textbf{\textcolor{black}{Question:}}
\emph{What is the man expressing that is so ironic?}
\begin{enumerate}[label=\textbf{\Alph*.}, nosep, leftmargin=*]
    \item \textbf{Desire to better the city.}
    \item Desire to sincerely apologize.
    \item Desire to keep his job.
\end{enumerate}
\vspace{3pt}
\textbf{\textcolor{black}{Question Type:}} Language Analysis
\end{memeexample}

% hobbit
\begin{memeexample}[colback=gray!2,colframe=light-gray,
  title={They're taking the hobbits to Isengard}, left=4pt,right=4pt,top=4pt,bottom=4pt]
% 1. Set font to small
\small
% 2. Set general text color to dark gray (70% black)
\color{black!70}

% --- Start of image insertion block ---
\noindent
\begin{minipage}[t]{0.5\linewidth}
    \vspace{0pt}
    \textbf{\textcolor{black}{Category:}} song \\
    \textbf{\textcolor{black}{Year:}} 2005 \\
    \textbf{\textcolor{black}{Language:}} English \\
\end{minipage}%
\hfill
\begin{minipage}[t]{0.35\linewidth}
    \vspace{0pt}
    \raggedleft
    % Placeholder image as requested
    \includegraphics[width=\linewidth, keepaspectratio]{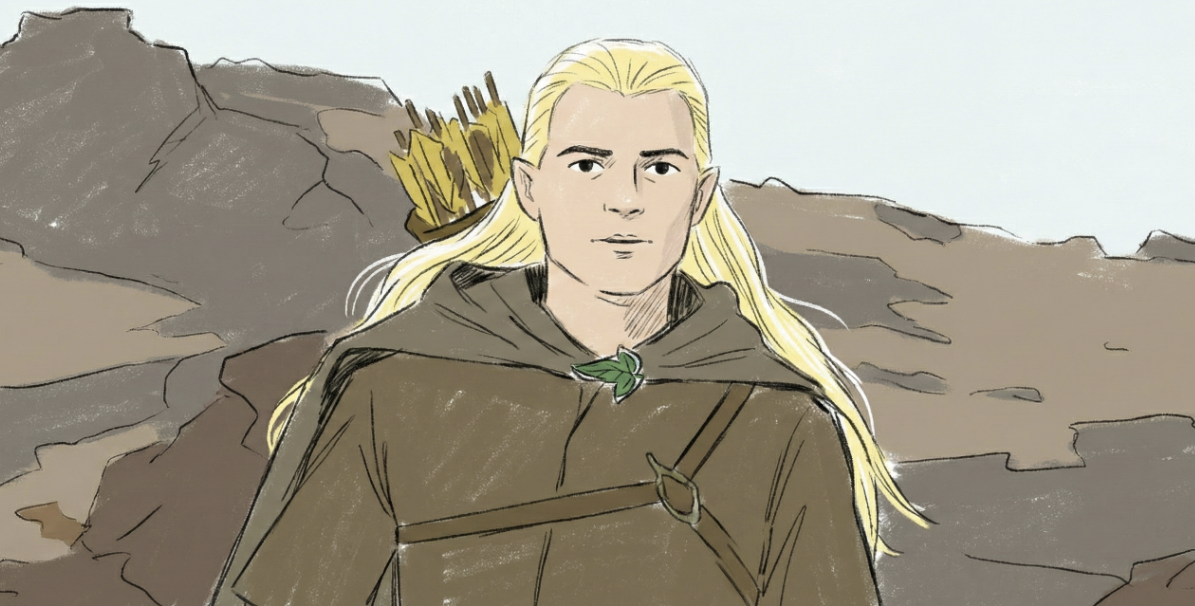}
\end{minipage}
\par\vspace{4pt}
% --- End of image insertion block ---
\begin{CJK}{UTF8}{min}
\textbf{\textcolor{black}{Transcription:}} They're taking the Hobbits to Isengard.
\end{CJK}\\
\textbf{\textcolor{black}{Summary:}}  A looping remix of Legolas from the Lord of the Rings repeatedly shouting “They're taking the hobbits to Isengard” edited to a catchy electronic beat.
 \\[3pt]
\textbf{\textcolor{black}{Usage:}} Used to humorously announce that someone is being taken somewhere, or as a nostalgic reference to early viral internet video culture.
 \\
\textbf{\textcolor{black}{Emotion:}} sarcastic/ironic, happy
 \\
\textbf{\textcolor{black}{Sensitivity:}} none
 \\
\textbf{\textcolor{black}{Question:}}
\emph{How does the popular meme version of this clip transform the mood of the original dramatic scene?}
\begin{enumerate}[label=\textbf{\Alph*.}, nosep, leftmargin=*]
    \item It makes the scene even more serious and tragic.
    \item It removes the audio and adds a laugh track.
    \item \textbf{It turns the scene into a catchy and repetitive electronic dance song.}
    \item It slows down the dialogue for emphasis.
\end{enumerate}
\textbf{\textcolor{black}{Question Type:}} Humor \& Popularity
\end{memeexample}

\end{multicols}

\end{document}